\documentclass[12pt]{spieman}  

\usepackage{amsmath,amsfonts,amssymb}
\usepackage{graphicx}
\usepackage[colorlinks=true, allcolors=blue]{hyperref}
\usepackage{textcomp}
\usepackage{gensymb}
\usepackage{multirow}
\usepackage{booktabs}
\usepackage{subfig}
\usepackage[left]{lineno}

\title{AIROPA IV: Validating Point Spread Function Reconstruction on Various Science Cases}

\author[a,*]{Sean K. Terry}
\author[a]{Jessica R. Lu}
\author[b]{Paolo Turri}
\author[c]{Anna Ciurlo}
\author[c]{Abhimat Gautam}
\author[c]{Tuan Do}
\author[c]{Michael P. Fitzgerald}
\author[c]{Andrea Ghez}
\author[c]{Matthew Hosek Jr.}
\author[d]{Gunther Witzel}

\affil[a]{Department of Astronomy, University of California, Berkeley, CA 94720, USA}
\affil[b]{Department of Physics \& Astronomy, University of British Columbia, Canada, V6T 1Z1}
\affil[c]{Division of Astronomy \& Astrophysics, University of California Los Angeles, CA 90095, USA}
\affil[d]{Max-Planck-Institut f\"{u}r Radioastronomie, Auf dem H\"{u}gel 69, Bonn, D-53121, Germany}

\begin{document} 
\pagecolor{white}
\maketitle

\begin{abstract}
We present an analysis of six independent on-sky datasets taken with the Keck-II/NIRC2 instrument. Using the off-axis point spread function (PSF) reconstruction software AIROPA, we extract stellar astrometry, photometry, and other fitting metrics in order to characterize the performance of this package. We test the effectiveness of AIROPA to reconstruct the PSF across the field of view in varying atmospheric conditions, number and location of PSF reference stars, stellar crowding and telescope position angle (PA). We compare the astrometric precision and fitting residuals between a static PSF model and a spatially varying PSF model that incorporates instrumental aberrations and atmospheric turbulence during exposures. Most of the fitting residuals we measure show little to no improvement in the variable-PSF mode over the single-PSF mode. For one of the data sets, we find photometric performance is significantly improved (by ${\sim}10\times$) by measuring the trend seen in photometry as a function of off-axis location. For nearly all other metrics we find comparable astrometric and photometric precision across both PSF modes, with a ${\sim}13$\% smaller astrometric uncertainty in variable-PSF mode in the best case. We largely confirm that the spatially variable PSF does not significantly improve the astrometric and other PSF fitting residuals over the static PSF for on-sky observations. We attribute this to unaccounted instrumental aberrations that are not characterized through afternoon adaptive optics (AO) bench calibrations.
\end{abstract}


{\noindent\footnotesize\textbf{*} \href{mailto:sean.terry@berkeley.edu}{sean.terry@berkeley.edu}}

\section{Introduction} \label{sec:intro}
With the next generation of adaptive optics (AO) instruments coming online soon, it is becoming increasingly important to properly characterize the spatial and temporal dependence of the point spread function (PSF) for data obtained with AO. The Keck-I and Keck-II AO systems have been used to deliver very high-resolution imaging for well over two decades, and have been continuously upgraded and fitted with newer generation hardware. The future of both Keck telescopes is filled with several promising next-generation updates \cite{wizinowich:2020a, bond:2020a}.
\\
\indent As a result of imperfect knowledge of the spatially varying (i.e. off-axis) PSF in these AO systems, very precise astrometry and photometry for a large majority of stellar sources in crowded fields (for example) has been limited. The Anisoplanatic and Instrumental Reconstruction of Off-axis PSFs for AO (AIROPA) is a suite of software packages that utilizes phase-diversity measurements, atmospheric profile data, and wave propagation through both turbulence and optical systems. With this knowledge, AIROPA generates a model of the field-dependent PSF for both natural guide star (NGS) and laser guide star (LGS) modes. The software functions under the assumption that every PSF that is extracted consists of a convolution of the on-axis PSF, the instrumental aberration, and the atmospheric anisoplanatism \cite{do:2018a}. Further descriptions of AIROPA and the sub-modules that it is built upon are given in \cite{witzel:2016a}.
\\
\indent For context we give a brief description of the input data needed for AIROPA. These data are used to generate the optical transfer function (OTF) grids (instrumental + atmospheric), on-axis and off-axis PSFs, and other files for the subsequent photometric and astrometric analyses. Instrumental aberration maps are generated by conducting fiber phase-diversity measurements at the detector plane for NIRC2. A grid of phase maps is generated, where each map is the result of the difference between the measured on-axis wavefront and off-axis wavefront across the 1024x1024 pixel field. Since the instrumental phase maps are mostly static \cite{Ciurlo:inprep}, the instrumental OTF can be read in from a pre-determined library of OTFs at various rotator angles. This significantly increases efficiency and reduces computation time for a given AIROPA analysis. The multi-aperture scintillation sensor (MASS) and differential image motion monitor (DIMM) are instruments on the summit of Mauna Kea that monitor the seeing and generate atmopheric profiles ($C_n^{2}$) at altitudes of 0.5, 1, 2, 4, 8, and 16 kilometers (km) above the summit. The algorithm that feeds this seeing information into AIROPA is called ARROYO \cite{britton:2006a}, and is built upon a set of C++ libraries. The AIROPA-generated OTF then represents the combination of the instrumental phase maps and atmospheric profile that are needed in order to construct the field-dependent PSF model. The PSF extraction and fitting is performed on each science image and final star lists are generated with photometry and astrometry for each detected source. Additionally, we rely on a fitting metric deemed the \textit{fraction of variance unexplained} (FVU), for determining how well the PSF model has fit the data in each image. For a schematic on the variable-PSF algorithm in AIROPA, we refer to Figure 1 in \cite{turri:2022}.
\\
\indent This study is the fourth in a series of papers detailing the AIROPA package; \cite{witzel:2016a} introduces AIROPA and gives an overview of the software structure. Characterization of the Keck-II/NIRC2 instrumental aberrations and AIROPA's usage of this aberration data is given in \cite{Ciurlo:inprep}, and \cite{turri:2022} perform tests of AIROPA on simulated and on-sky Galactic Center (GC) images. In this paper we focus on expanding the on-sky tests of AIROPA. The paper is organized as follows: Section \ref{sec:gc-data} describes the analysis of a very crowded field through three independent GC datasets. In Section \ref{sec:ogle-data}, we describe a less-crowded field for a typical high-resolution gravitational microlensing science case. In Section \ref{sec:m53-data} we detail the observations of globular cluster Messier 53 (M 53) taken at various position angles (PA). In Section \ref{sec:results} we compare the photometric, astrometric, and FVU results for each science case. Finally, we give a discussion and conclude the paper in Section \ref{sec:conclusion}.


\section{Observations} \label{sec:observations}
All datasets presented in this work were acquired on Keck-II with the Near-Infrared Camera 2 (NIRC2) narrow camera instrument in laser guide star adaptive optics (LGSAO) mode and with the K$_\textrm{p}$ filter ($\lambda_{c} = 2.12\,\,\mu$m). The pixel scale for the NIRC2 narrow camera is 9.952 mas/pix \cite{service:2016a}, and all the data were taken between May 2015 and August 2017.

\begin{figure}[!h]
 \makebox[\textwidth][c]{
 \includegraphics[width=1.2\textwidth]{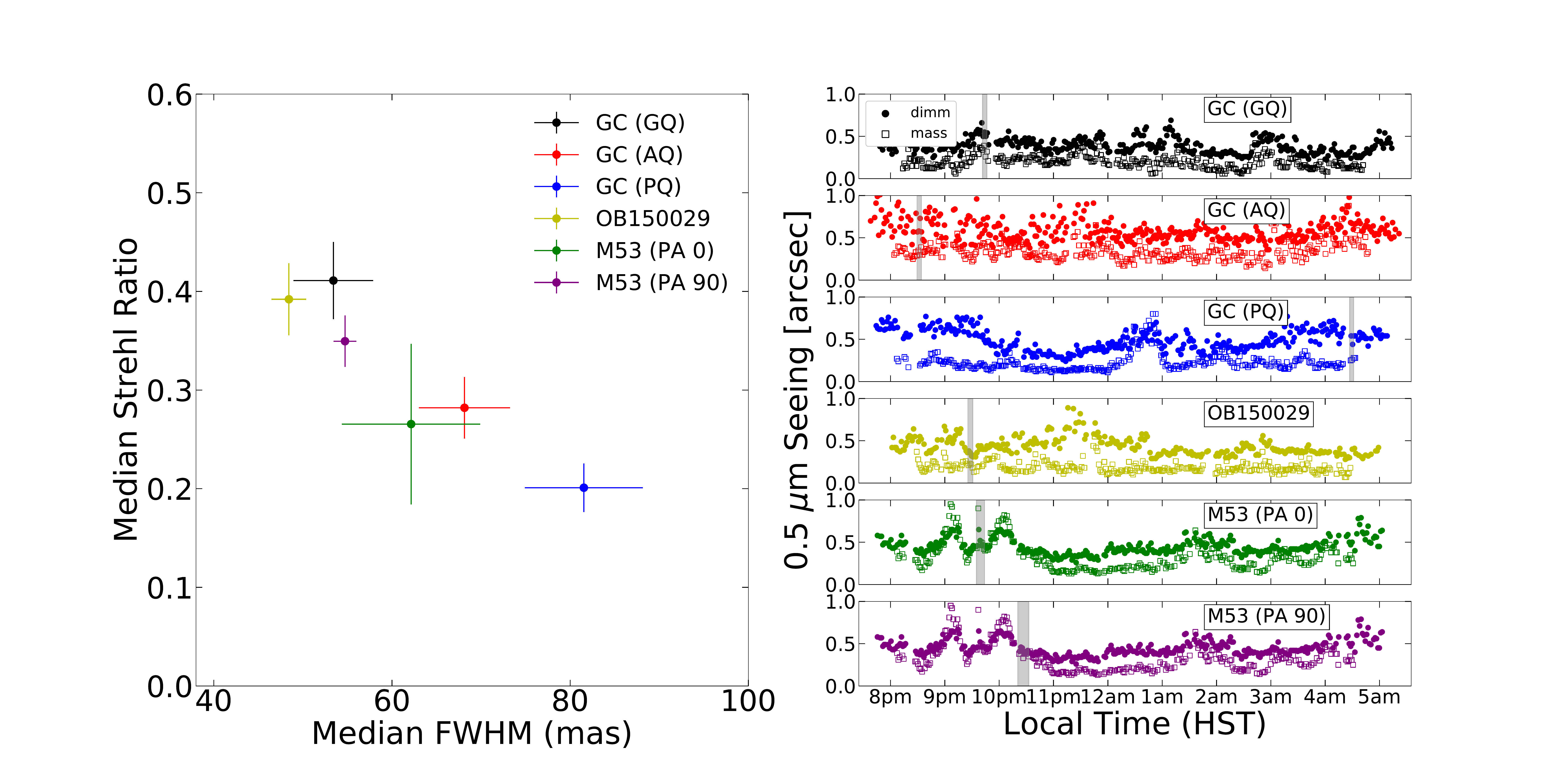}
 }
 \caption{\footnotesize \textit{Left}: Median SR and FWHM values for all datasets analyzed in this work. \textit{Right}: 0.5\,\,$\mu$m DIMM/MASS seeing profiles for each dataset, respectively. Grey shaded region represents the time of observation for each set. \label{fig:dimmmass_datasets}}
\end{figure}

\indent A total of six cleaned frames from each dataset were reduced identically with a NIRC2 pipeline that corrects for differential atmospheric refraction (DAR), bad pixels, cosmic rays, and other undesirable effects \cite{ghez:2008a, lu:2008a} and applies geometric distortion corrections \cite{lu:2008a, service:2016a}. North is up and east is left in all observations, with one exception for the PA $=$ 90$^\circ$ observations where east is up and south is left. Figure \ref{fig:dimmmass_datasets} shows the median Strehl ratio (SR) and full width half maximum (FWHM) values and standard deviations for all datasets analyzed in this work. The right panel of Figure \ref{fig:dimmmass_datasets} shows the atmosphere profile information from the DIMM/MASS instruments for each dataset, with grey shaded regions representing the time of observation considered in this analysis. The observation timestamps span 8:30 pm to 4:30 am local Hawaiian Standard Time (HST). Table \ref{tab:fields-metrics} shows the quality metrics: median SR, FWHM, and root-mean-square deviation of the wavefront error (RMS WFE). The table also gives observation dates and total exposure times for all datasets.
\\
\indent There are many reasons to include a wide range of variable condition data across different stellar fields on-sky, which include testing the effects of varying DIMM/MASS profiles on PSF extraction (Section \ref{sec:gc-data}), testing the fitting precision on datasets with more (or less) PSF reference stars in the field (i.e. crowded or sparse fields, Sections \ref{sec:ogle-data} and \ref{sec:m53-data}), and determining the reliability of goodness-of-fit and other PSF fitting residuals for extracted sources (Sections \ref{sec:results-gc}, \ref{sec:results-ob150029}).

\begin{table}[!h]
\centering
\caption{Observational data presented in Sections \ref{sec:gc-data}, \ref{sec:ogle-data}, and \ref{sec:m53-data}} \label{tab:fields-metrics}
\begin{tabular}{lccccc}
\hline
        Field &  Date [UT] &  Strehl Ratio &  FWHM [mas] &  RMS WFE [nm] & Exp. Time [s] \\\hline\hline
        GC (GQ) &   2017 Aug 11 & $0.41 \pm 0.039$ & $53 \pm 4.5$ & $320 \pm 18$ & 168\\
        GC (AQ) &   2017 Aug 23 & $0.29 \pm 0.024$ & $66 \pm 3.5$ & $380 \pm 13$ & 168\\
        GC (PQ) &   2016 May 03 & $0.21 \pm 0.025$ & $82 \pm 6.6$ & $430 \pm 18$ & 168\\
        OB150029 &    2016 Jul 14 &  $0.39 \pm 0.037$ & $48 \pm 1.9$ & $330 \pm 17$ & 180\\
        M53 (PA 0) &   2015 May 05 &  $0.27 \pm 0.081$ & $62 \pm 7.8$ & $390 \pm 44$ & 300\\
        M53 (PA 90) &   2015 May 05 & $0.35 \pm 0.026$ & $55 \pm 1.3$ & $350 \pm 12$ & 300\\\hline
\end{tabular}
\end{table}

\subsection{Comparing Atmospheric Conditions in a Very Crowded Field: The Galactic Center} \label{sec:gc-data}
We expand upon the initial on-sky GC testing of \cite{turri:2022} by including ``good quality," ``average quality," and ``poor quality" datasets (hereafter GQ, AQ, and PQ respectively) for further AIROPA validations on the GC. The GC case was used as the main science driver for the original development of AIROPA. This case is ideal since there already exists a rich high-resolution dataset spanning several decades, dozens of reference stars for PSF modeling, and a very bright, uniform tip/tilt (TT) star that is ${\sim}13$ arcseconds off-axis. The longest-running high-resolution study of the stellar population immediately surrounding Sgr A$^{*}$ is being conducted by the Galactic Center Orbit Initiative\footnote[1]{\url{https://galacticcenter.astro.ucla.edu/}}.
This work has led to a deep and well-understood knowledge \cite{ghez:2005b, ghez:2008a, lu:2008a, do:2019a, gautam:2019a} of the environment immediately surrounding the central supermassive black hole.

\begin{figure}[!h]
 \makebox[\textwidth][c]{
 \hspace*{-0.5cm}\includegraphics[width=1.15\textwidth]{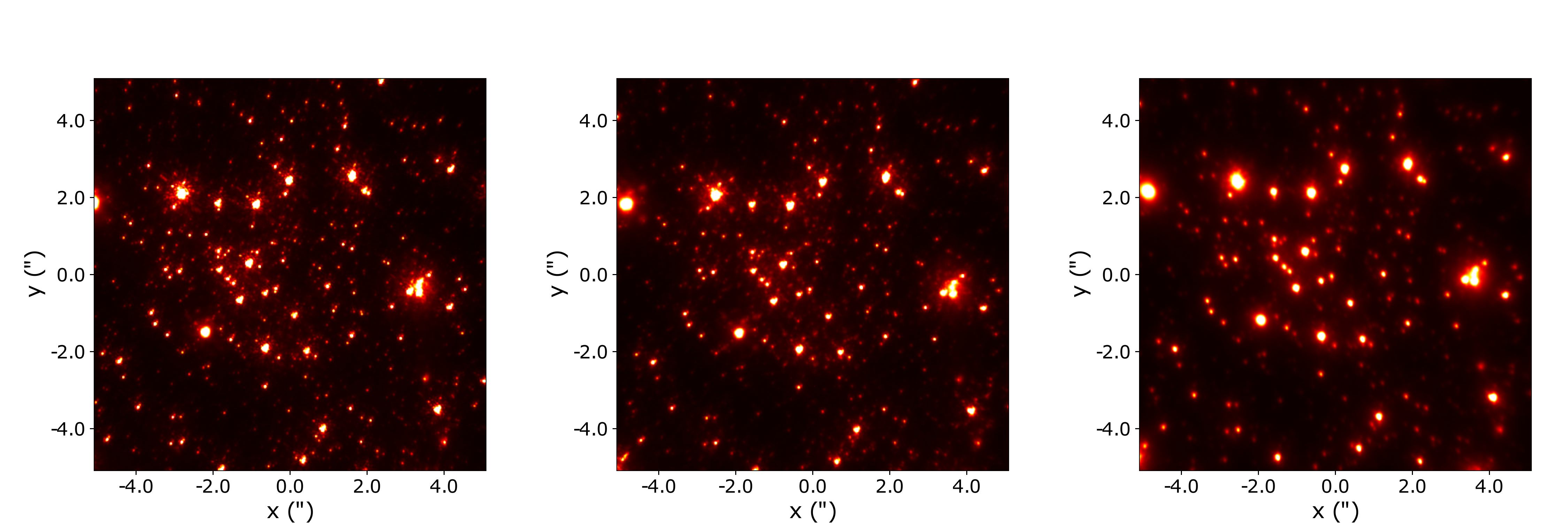}
 }
 \caption{\footnotesize \textit{Left}: Good-quality Galactic Center frame. \textit{Middle}: Average-quality Galactic Center frame. \textit{Right}: Poor-quality Galactic Center frame. All images have the same color scale, the axes represent on-sky separation (in arcsecond) from the LGS position. North is up and East is left in all frames. \label{fig:gc_images}}
\end{figure}

\begin{figure}
 \makebox[\textwidth][c]{
 \hspace*{0.5cm}\includegraphics[width=1.2\textwidth]{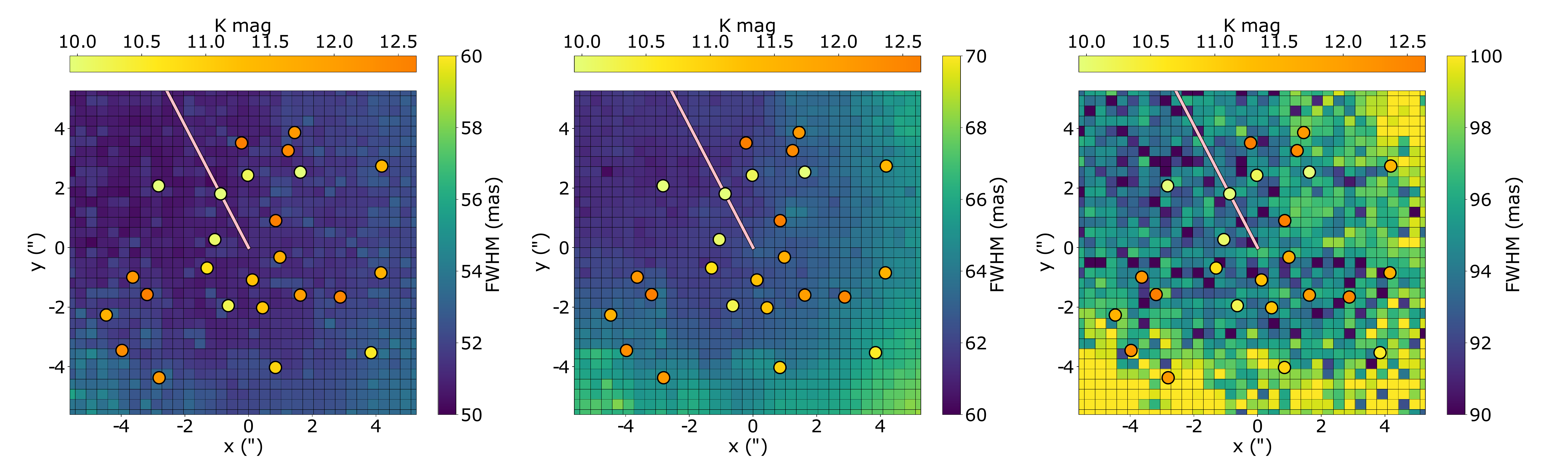}
 }
 \caption{\footnotesize Grid of FWHM values from AIROPA variable-PSF mode on the GQ (\textit{left}), AQ (\textit{middle}) and PQ (\textit{right}) datasets. Colored circle data points correspond to locations and K-band magnitudes of primary PSF reference stars. A solid line connects the on-aixs LGS position to the off-axis TT guide star position. \label{fig:fwhm_grids}}
\end{figure}

\indent We rank the three GC datasets, GQ, AQ, and PQ based on the historical quality of all GC epochs taken with NIRC2 since 2004 \cite{jia:2019a}. The GQ data were taken on 2017-08-11, The AQ data were taken $\sim$two weeks after the GQ data, on 2017-08-23, and the PQ data were taken on 2016-05-03. For all GC datasets, each frame was composed of 10 coadded exposures at 2.8 s per coadd for a total exposure time of 28 s per frame. One note about the PQ dataset --- there does not exist phase-diversity calibration data from 2016, therefore the 2017 instrumental phase maps were used to substitute for the 2016 phase maps. This is not an issue, as \cite{Ciurlo:inprep} show that the phase maps can remain stable (within ${\sim}59$ nm) across multiple years, and any potential difference between 2016 and 2017 phase-diversity is likely small enough for an analysis of the kind presented in this work. Figure \ref{fig:gc_images} shows one frame from each of the three epochs, with the spatial scale given as the separation (in arcseconds) from the on-axis position (i.e. location of the LGS). We note that there are two other axes of importance; the direction of the TT star (given by the solid line in Figures \ref{fig:fwhm_grids}, \ref{fig:ob150029}, and \ref{fig:m53_image_grid}), and the position of the image sharpening (typically offset from the center of the image). Figure \ref{fig:fwhm_grids} shows the grid of FWHM values measured by AIROPA variable-PSF mode for each frame, plotted over the field of view to give a visualization of the field-dependent PSF. The FWHM grid values are calculated for every PSF generated by AIROPA in a PSF grid file. The AIROPA grid files for the fields analyzed in this work use a partition size (i.e. step width of the PSF grid) of 102 pix. 


 
\subsection{A Less Crowded Field: OGLE-2015-BLG-0029} \label{sec:ogle-data}
The Optical Gravitational Lensing Experiment (OGLE) survey \cite{udalski:1992a} conducts wide-field visible and near-IR imaging of nearly the entire Galactic Bulge region at high cadence and detects over 1000 microlensing events every season. Most ground-based imaging data from current microlensing surveys are focused on the regions $3\degree > l > 357\degree$ and $-2.5\degree < b < 2.5\degree$. The first non-GC data analyzed in this work is OGLE-2015-BLG-0029 (hereafter OB150029), located at RA $=$ 17:59:46.60, dec $=$ -28:38:41.80 (J2000) and Galactic coordinates ($l,b = (1.828\degree, -2.523\degree$)). This target has been regularly monitored with NIRC2 since 2015 as part of a project to study isolated stellar-mass black hole candidates \cite{lu:inprep}.
\\
\begin{figure}[!h]
 \makebox[\textwidth][c]{
 \includegraphics[width=1.1\textwidth]{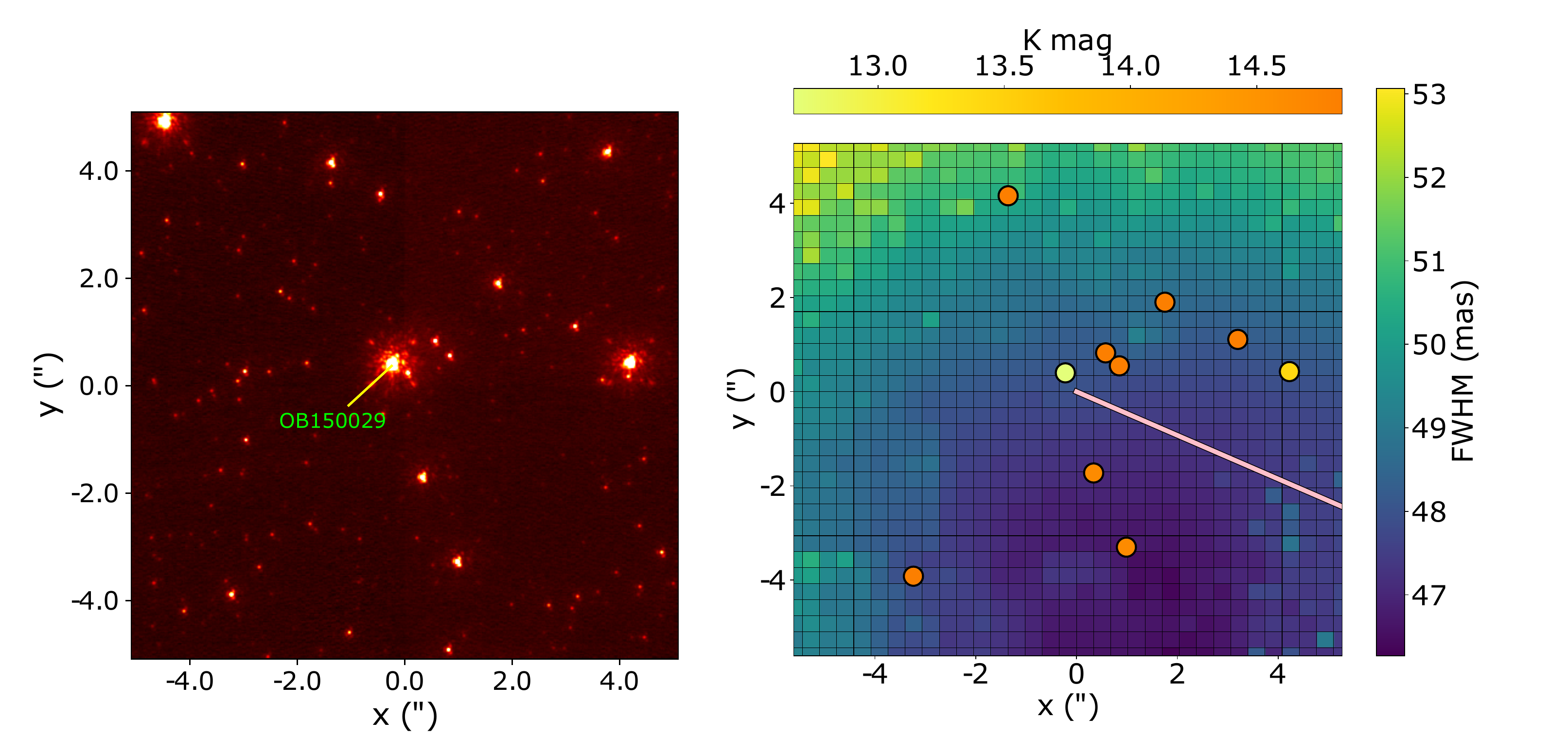}
 }
 \caption{\footnotesize \textit{Left}: NIRC2 frame of the OB150029 field taken on 2016-07-14 with the Kp filter. \textit{Right}: FWHM grid from AIROPA variable-PSF mode with selected PSF reference stars colored by magnitude and solid line connecting the on-axis position to the TT guide star.} \label{fig:ob150029}
\end{figure}
\indent The OB150029 observations were taken on 2016-07-14 in LGSAO mode with the Kp filter. Each of the frames consist of six co-added images, each with five second integration time, for a total integration time of 30 seconds per frame. The airmass during the six-frame observing window was $\sim$1.65, and the median SR, FWHM, and RMS WFE values for the six frames are given in Table \ref{tab:fields-metrics}, as well as the DIMM/MASS data given on the right-hand panel of Figure \ref{fig:dimmmass_datasets}. This is the second-best dataset (behind GQ) in terms of quality metrics like the median SR and FWHM and their variances across all six frames. Similar to the PQ dataset, there are no 2016 phase maps available for NIRC2, therefore the 2017 maps were substituted. Figure \ref{fig:ob150029} shows a NIRC2 image of the field, which is clearly less crowded than the GC, particularly in the bright star regime. 
\\
\indent There are a total of 10 PSF reference stars in this field that were used for AIROPA, compared to a total of 25 PSF reference stars used for the GC analysis described in Section \ref{sec:gc-data}. The total number and spatial location of selected PSF reference stars is important for constructing accurate PSF models, and the 10 stars chosen for the OB150029 field include the microlensing target itself (i.e. brightest star in the field), as well as stars within 1.5 mag of the target, and separations of $\pm4$ arcseconds from the target. The TT guide star used for the observations has an R mag ${\sim}$15.1 and separation of ${\sim}$13.8 arcseconds to the south-west of the target, as indicated by the solid line in the right panel of Figure \ref{fig:ob150029}.


\subsection{Comparing Different Position Angles: Globular Cluster M53} \label{sec:m53-data}
The globular cluster M53 (NGC 5024) located at RA $=$ 13:12:54.51, dec $=$ 18:10:13.95 (J2000) was observed with the NIRC2 narrow camera Kp filter on 2015-05-04 as part of a project to characterize the NIRC2 geometric distortion \cite{service:2016a}. This was accomplished by comparing the NIRC2 stellar positions to precise astrometry from \textit{Hubble Space Telescope} (HST) \textit{Advanced Camera for Surveys} (ACS) observations \cite{anderson:2008a}. The NIRC2 camera and AO system was realigned in April 2015, and observations of M53 before and after this realignment show an increase in the average geometric distortion from $\sim$0.5 mas pre-realignment to $\sim$1.1 mas post-realignment.
\\
\begin{figure}[!h]
 \makebox[\textwidth][c]{
 \includegraphics[width=1.0\textwidth]{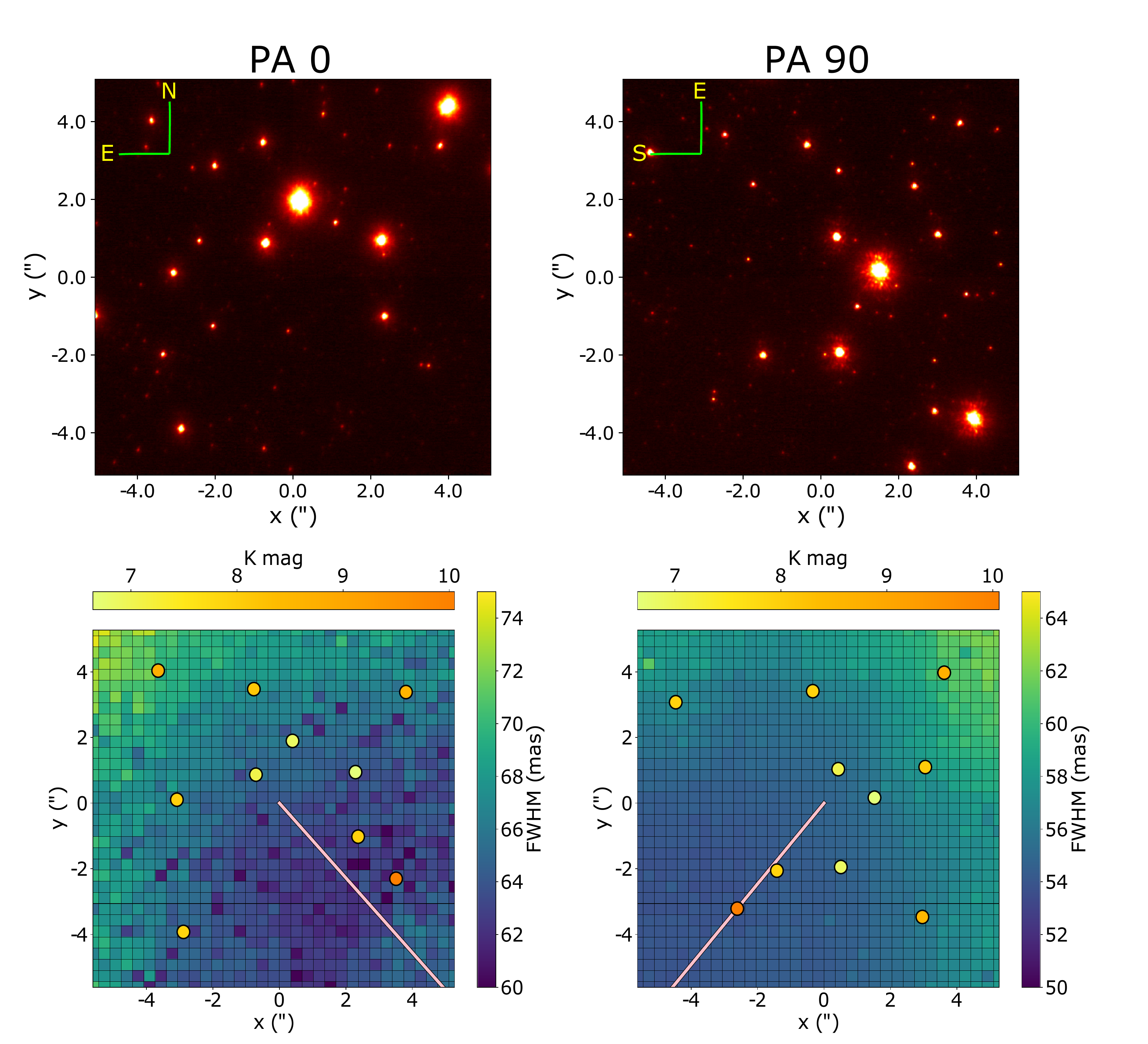}
 }
 \caption{\footnotesize \textit{Left Column}: M53 NIRC2 frame taken at PA = 0$^\circ$, with accompanying FWHM grid generated by AIROPA. \textit{Right Column}: Same as left, but for PA = 90$^\circ$. The solid line connects the on-axis LGS position to the TT guide star. The color scale is identical for both images in the top row.} \label{fig:m53_image_grid}
\end{figure}
\indent Importantly for the current study, the M53 NIRC2 observations were taken at two different position angles (PA); 0$\degree$ and 90$\degree$. These observations present a unique dataset for us to compare the astrometric accuracy between single-PSF and variable-PSF modes in AIROPA as a function of PA. The previous study of \cite{service:2016a} assumed a static PSF to derive the geometric distortion solution, which implies that using a variable-PSF on NIRC2 datasets will likely show residual distortion that was not modeled out by the static PSF used to derive the solution. Depending on the significance of this effect, some fraction of the difference in astrometric, photometric, or FVU results between the PSF modes may be attributed to this (Figures \ref{fig:m53_PA_compare}, \ref{fig:m53_PA_compare_hst}, and \ref{fig:m53_PA_compare_gaia}). Due to the gradient in the NIRC2 distortion solution, the local pixel scale changes across the detector. This change, however small, is not detected in any PSF modeling because the aberration maps that are used are tip-tilt removed. This is very likely only a minor effect. As our intention in this paper is not to fully derive a new geometric distortion solution (with single and/or variable-PSF), we select only a subset of M53 NIRC2 images to remain consistent with the number of GC and microlensing frames used in previous sections. While it may prove beneficial to fully derive a new independent variable-PSF distortion solution for NIRC2, this is beyond the scope of the current paper. As shown in \cite{turri:2022} and Figure \ref{fig:gc-gq-astrom}, the astrometric differences between single-PSF mode and variable-PSF mode become more severe at larger radii from the central on-axis position. This is predominantly caused by the difference in PSF shape between the spatially varying model and the static model, particularly closer to the sides and edges of the NIRC2 frame.
\\
\indent The M53 dataset includes six frames taken at a position angle of 0$^\circ$ (north up, east left), and six frames taken at a position angle of 90$^\circ$ (east up, south left). The PA $=0^\circ$ dataset consists of four frames taken at the central pointing, and two frames taken after a dither of $\Delta_x = -0.14", \Delta_y = 0.16"$. The PA $=90^\circ$ dataset consists of four frames taken at the central pointing, and two frames taken after a dither of $\Delta_x = +0.14", \Delta_y = -0.19"$. The total exposure time is 50 seconds per frame for both PAs. Figure \ref{fig:m53_image_grid} shows one frame taken at PA 0$^\circ$ and the accompanying grid of FWHM values from AIROPA variable-PSF mode, as well as a PA 90$^\circ$ frame and FWHM grid. The same 10 PSF reference stars were used for each PA, as well as the same TT guide star, which has magnitude R ${\sim}$ 13.5 located 24 arcseconds to the south-west (solid line in the lower panels of Figure \ref{fig:m53_image_grid}).


\section{Results} \label{sec:results}
We present the AIROPA single-PSF and variable-PSF results for each field analyzed. As mentioned earlier, to maintain consistency we selected a subset of six consecutive frames from each dataset and used the averaged astrometry, photometry, and FVU in our final analysis and comparisons. All of the selected GC data (as well as the microlensing data) have exactly one dither between the first three frames and last three frames, while the M53 data have exactly one dither between the second and third frames (i.e. first two frames at one dither position, last four frames at a subsequent dither position).
\\
\begin{figure}[!h]
 \makebox[\textwidth][c]{
 \includegraphics[width=0.95\textwidth]{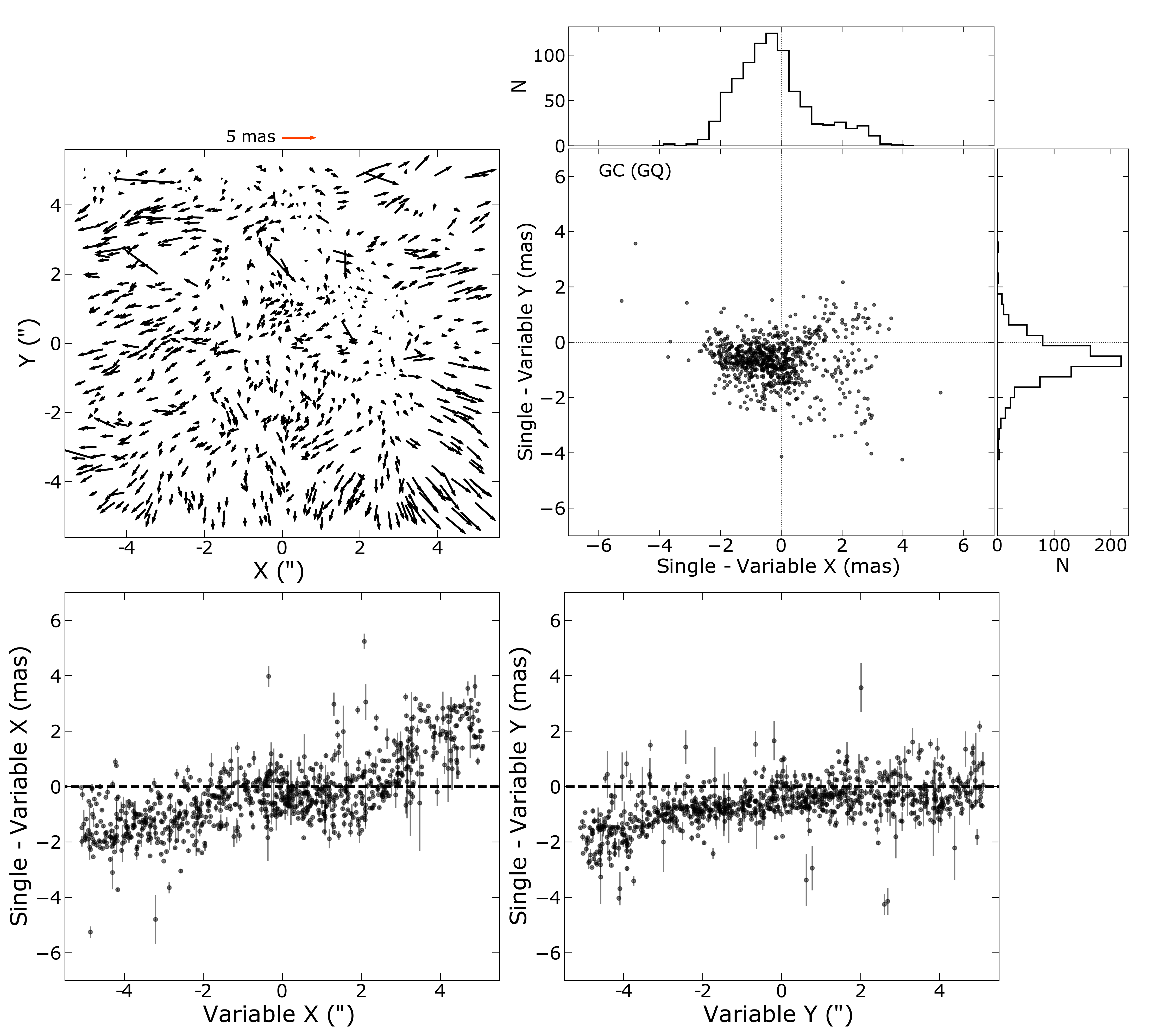}
 }
 \caption{\footnotesize \textit{Top row}: 2D positional difference in single vs. variable PSF modes. Quiver plot shows positional differences as a function of field location. \textit{Bottom row}: 1D positional differences with errors derived from the RMS deviation of six positional measurements.} \label{fig:gc-gq-astrom}
\end{figure}
\indent There are three primary metrics that we used to assess the performance of each PSF mode in AIROPA, these are the photometric uncertainty, astrometric uncertainty, and FVU of the residual image at the given position. We do expect the varying conditions, data quality, PSF reference stars, and stellar crowding to impact AIROPA's effectiveness to reconstruct the PSF. Along with instrumental aberrations and atmospheric anisoplanatism, the combination of these factors (on the instrument and on-sky) work to effectively increase the overall noise present in the reconstructed PSFs, stellar profiles, and extracted photometry, astrometry, and FVU. For each dataset, a relative comparison of these metrics between the PSF modes will reveal any differences in the overall PSF-R performance or results for each mode.

\begin{table}[!h]
\caption{Photometric, astrometric differences in single and variable PSF modes for bright sources $m \leq 13$}
\setlength{\tabcolsep}{12.0pt}
\begin{center}
\begin{tabular}{lccccc}
    \hline\hline
    {} & {} & \multicolumn{2}{c}{$\Delta m_{(S-V)}$ [mag]} & \multicolumn{2}{c}{$\Delta r_{(S-V)}$ [mas]}\\
    \cmidrule(lr){3-4}\cmidrule(l){5-6}
        Field & $N_{\textrm{bright}}$ & $r \leq 3$\textquotesingle\textquotesingle & $r > 3$\textquotesingle\textquotesingle & $r \leq 3$\textquotesingle\textquotesingle & $r > 3$\textquotesingle\textquotesingle\\
        \hline
        GC (GQ) & 140 & $-0.05 \pm 0.02$ & $0.01 \pm 0.03$ & $0.08 \pm 0.52$ & $1.01 \pm 0.75$\\
        GC (AQ) & 127 & $-0.03 \pm 0.03$ & $0.02 \pm 0.05$ & $0.08 \pm 0.74$ & $1.01 \pm 0.98$\\
        GC (PQ) & 112 & $-0.02 \pm 0.02$ & $0.03 \pm 0.04$ & $0.07 \pm 0.84$ & $0.76 \pm 1.25$\\
        OB150029 & 19 &  $0.03 \pm 0.02$ & $0.09 \pm 0.04$ & $0.12 \pm 0.39$ & $1.02 \pm 0.56$\\
        M53 (PA 0) & 32 &  $-0.01 \pm 0.04$ & $0.04 \pm 0.05$ & $0.15 \pm 0.68$ & $0.95 \pm 0.96$\\
        M53 (PA 90) & 28 & $0.02 \pm 0.03$ & $0.10 \pm 0.05$ & $0.12 \pm 0.68$ & $0.55 \pm 1.40$\\\hline
\end{tabular}
\end{center}
\label{tab:PhotAstromDiff-results}
\end{table}

\begin{figure}[!h]
  \subfloat{\includegraphics[width=0.333\textwidth]{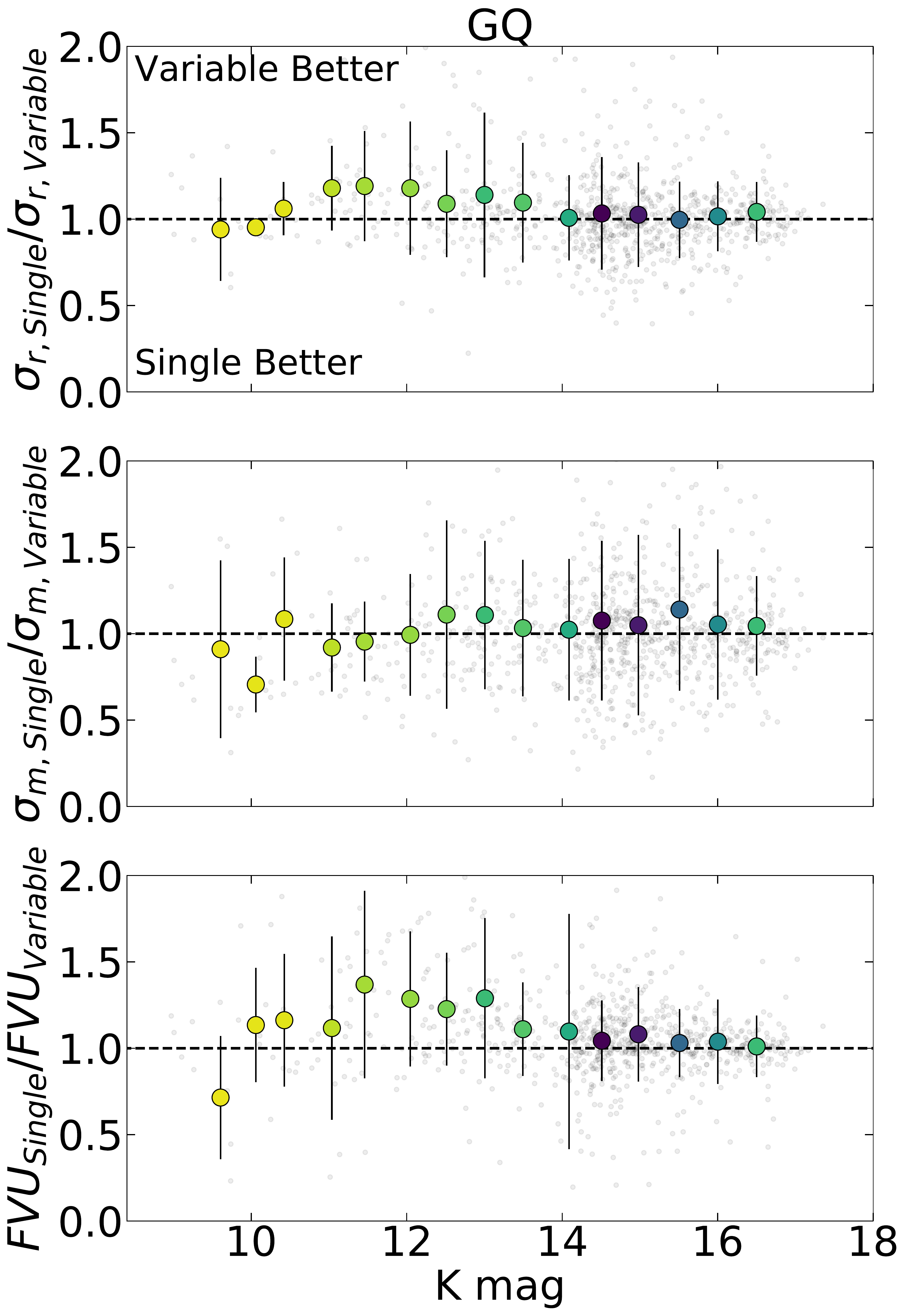}}
  \subfloat{\includegraphics[width=0.289\textwidth]{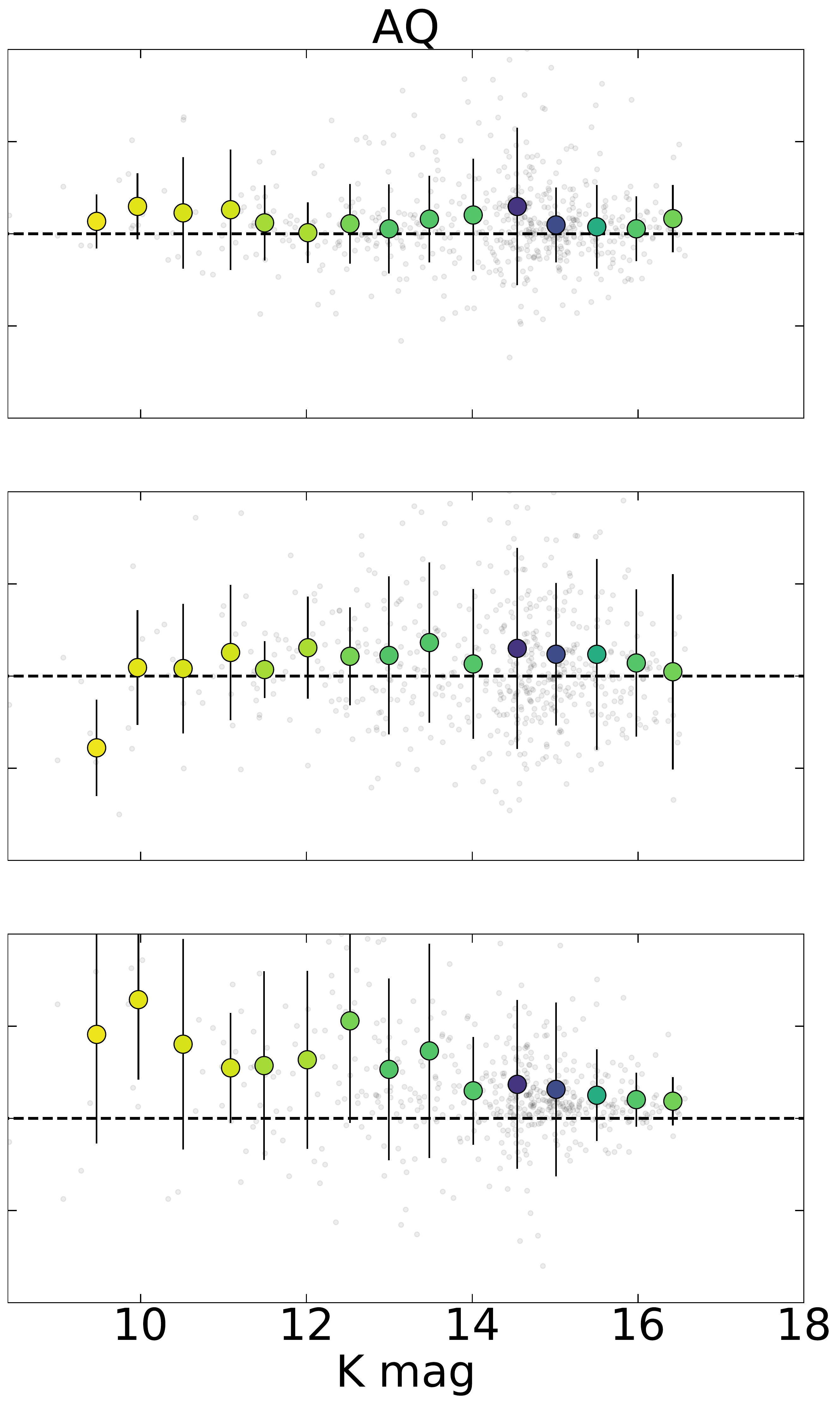}}
   \subfloat{\includegraphics[width=0.345\textwidth]{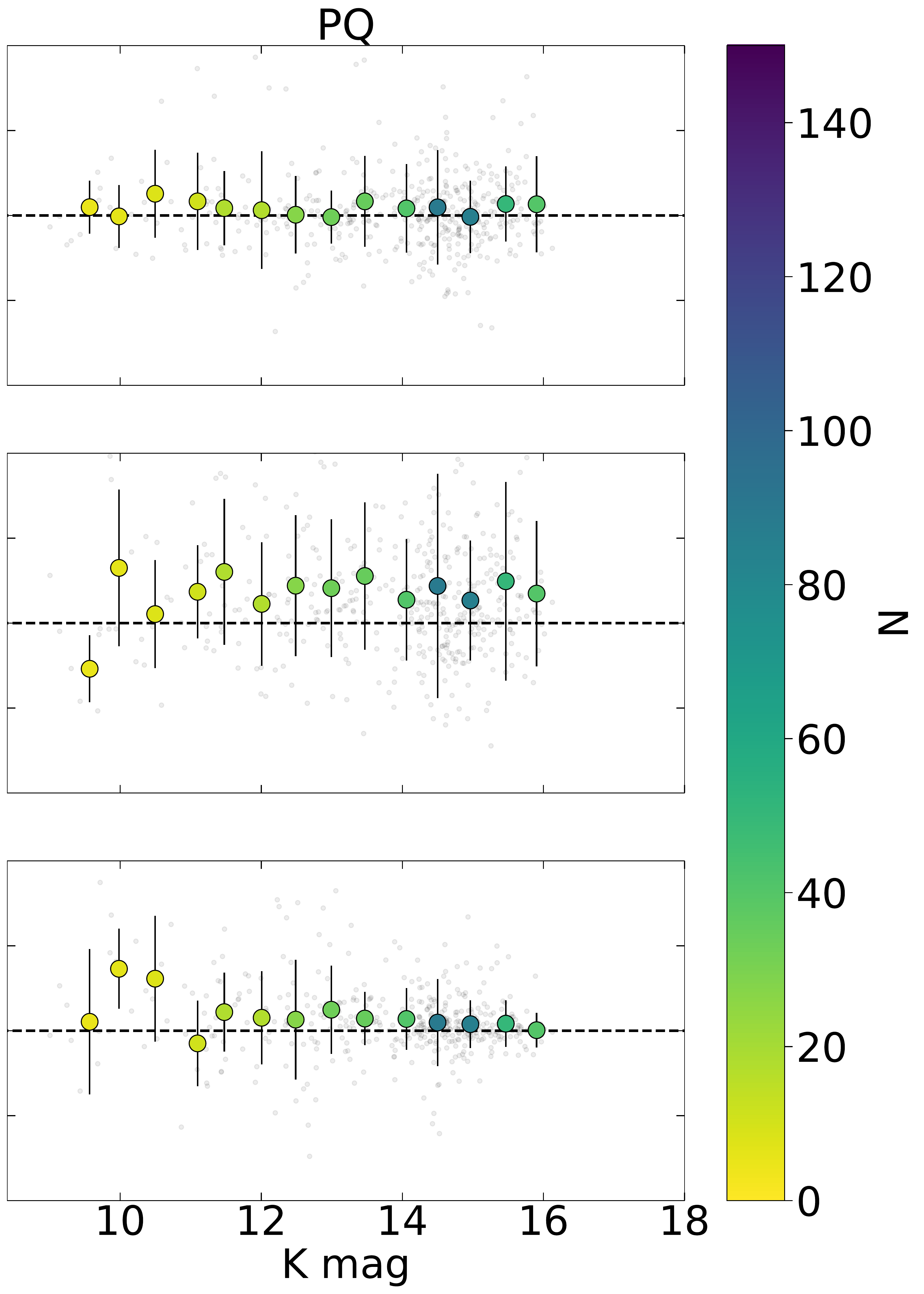}}
  \caption{\footnotesize Ratio of astrometric errors, photometric errors, and FVU values for the GC datasets; GQ (\textit{left column}), AQ (\textit{middle column}), and PQ (\textit{right column}). Values larger than one indicate the variable-PSF mode resulted in smaller astrometric/photometric residuals and smaller FVU, respectively. All stars are plotted in faint grey. Blue data points represent 0.5 mag bins, with error bars representing the 1$\sigma$ standard deviation of the ratios for stars in each bin.} \label{fig:gc-astrom}
\end{figure}

\indent In addition to the three primary metrics mentioned above, we also compare the M53 astrometry and photometry from each PSF mode to several external catalogs. We make these comparisons in an attempt to determine which PSF mode gives results that are more consistent with an external reference. As shown in Figures \ref{fig:gc-gq-astrom} and \ref{fig:gc-astrom}, and in \cite{turri:2022}, there is a clear relative difference in astrometry and photometry between the two PSF modes in AIROPA. However, information on relative differences alone does not tell us which PSF mode produces results that are closest to truth. This requires comparisons like those described in the subsequent M53 results sections.
\\
\indent The various astrometry, photometry, FVU, and source color results are presented in several tables throughout the results section. Table \ref{tab:PhotAstromDiff-results} reports the difference in measured magnitudes and source positions in the two PSF modes. The stars and subsequent measurements are grouped into two radii; $r \leq 3$\textquotesingle\textquotesingle, where the anisoplanatic effect on the PSFs should have a smaller impact, and $r > 3$\textquotesingle\textquotesingle, where anisoplanatism significantly affects the shape of PSFs. Ideally, a spatially varying reconstructed PSF (i.e. AIROPA variable-PSF mode) properly accounts for the anisoplanatic effect by shaping the modeled PSF according to the instrumental and atmospheric components of the OTF at the larger off-axis locations. Table \ref{tab:PhotAstromFVU-results} gives the average photometric and astrometric errors, and average FVU results for all datasets analyzed in this work. The table includes results for all detected stars in each set, as well as results for the brightest stars ($m_K < 13$), as the PSF residuals at the brightest magnitudes are affected primarily by PSF systematic errors rather than noise \cite{turri:2022}. This effect can be seen in the FVU comparisons for bright stars in each field (Figures \ref{fig:gc-astrom}, \ref{fig:ob-m53-astromfvu-ratios} and the right columns in Table \ref{tab:PhotAstromFVU-results}).

\subsection{Galactic Center} \label{sec:results-gc}

\indent For the very crowded GC datasets, identical reductions and analyses were performed across each of the six-frame sets. The same 25 PSF reference stars were used in each analysis and the astrometric, photometric, and FVU results are given in the following three sub-sections. Note that for each of the three datasets, we impose a requirement that in order for a star to be included in the analysis, it must be detected in at least five of the six cleaned frames.

\subsubsection{GC: Good Quality (GQ)} \label{sec:results-gc-GQ}
\noindent The GQ dataset contains the largest number of stars detected of all data analyzed in this paper. A total of 840 stars were detected in at least five of the six GQ frames. Figure \ref{fig:gc-gq-astrom} shows the averaged difference (from the six frames) between the measured star positions in single-PSF and variable-PSF modes. The astrometric differences are largest at off-axis locations farthest from the frame center, in part due to the difference in PSF shape between the static and the spatially variable PSF models. The top right panel of Figure \ref{fig:gc-gq-astrom} shows the field position for each detected star and the corresponding 2D position difference on the detector plane. The 1D differences are shown in the bottom panels of Figure \ref{fig:gc-gq-astrom}. The mean astrometric difference in each direction across the detector for bright stars inside and outside of a 3\textquotesingle\textquotesingle\,off-axis radius is given in Table \ref{tab:PhotAstromDiff-results}, along with the mean magnitude differences.
\\
\indent The average photometric and astrometric precision for all GQ stars in both PSF modes is $\sigma_{{m}}{\sim}0.04$ mag and $\sigma_{{r}}{\sim}0.81$ mas, respectively. Further, the average FVU metric is very similar for the variable-PSF mode ($4.67\times10^{-2}$) and single-PSF mode ($4.70\times10^{-2}$). There may be evidence of some improvement in the astrometric errors and FVUs for only the brightest stars in the field (i.e. left column of Figure \ref{fig:gc-astrom}), but any possible improvement here is significantly less than what has been measured on simulation tests under similar conditions \cite{turri:2022}.

\subsubsection{GC: Average Quality (AQ)} \label{sec:results-gc-AQ}
The total number of stars detected across the AQ frames is 576, approximately 30\% less than the GQ dataset. As expected, the average uncertainties are marginally larger than those from the GQ dataset. When comparing the results from both PSF modes, we measure marginally smaller uncertainties for the variable-PSF mode over the single-PFS mode. While these results imply that variable-PSF is performing better than single-PSF mode, we note again that these improvements are still well below what has been shown through simulated data tests.
\\
\indent The average photometric precision for the AQ stars is $\sigma_{{m}}{\sim}0.038$ mag for variable-PSF mode and $\sigma_{{m}}{\sim}0.040$ mag for single-PSF mode. Further, the measured astrometric precision is $\sigma_{{r}}{\sim}1.07$ mas for variable-PSF mode and $\sigma_{{r}}{\sim}1.11$ mas for single-PSF mode. Finally, the average FVU metric is smaller (${\sim}10$\%) for the variable-PSF mode ($4.52\times10^{-2}$) compared to single-PSF mode ($5.01\times10^{-2}$).

\subsubsection{GC: Poor Quality (PQ)} \label{sec:results-gc-PQ}
As expected the PQ dataset had the smallest number of detected stars, at 445, this is ${\sim}23$\% less than the AQ dataset and ${\sim}47$\% less than the GQ dataset. As a reminder, Table \ref{tab:fields-metrics} gives the median PSF FWHM for the PQ dataset, $82 \pm 6.6$ mas. This is approximately 30 mas larger than the GQ PSF and approximately 16 mas larger than the AQ PSF.
\\
\indent Similarly to the previous GC datasets, the PQ results show a marginal improvement for the variable-PSF mode compared to single-PSF mode. The astrometric precision for variable-PSF mode ($\sigma_{{r}}{\sim}1.32$ mas) is just 2\% less than single-PSF mode ($\sigma_{{r}}{\sim}1.35$ mas). A better improvement is seen in the photometric precision comparison; variable-PSF mode gives a precision of $\sigma_{{m}}{\sim}0.042$ mag which is ${\sim}10$\% smaller than the single-PSF mode precision ($\sigma_{{m}}{\sim}0.047$). Finally, the average FVU measurement across all detected stars is once again quite similar between the two modes; variable-PSF gives an average FVU of $4.51\times10^{-2}$ while single-PSF gives an average FVU of $4.64\times10^{-2}$.

\subsection{OB150029} \label{sec:results-ob150029}

\indent The microlensing target OB150029 is located in the least-crowded stellar field analyzed in this work, with a total of 63 stars cross-matched in at least five out of six frames. For this data, 10 reference stars were used for the initial PSF model. Coincidentally, this is the same amount of reference stars as the M53 dataset, and is 60\% fewer reference stars than the GC datasets. We note this dataset has the second-highest measured SR, and smallest FWHM of all datasets analyzed (Table \ref{tab:fields-metrics}). Additionally, as can be seen in the right panel of Figure \ref{fig:ob150029}, this dataset has the smallest variance in FWHM as measured across the field (${\sim}6$ mas). 
\\
\begin{figure}[!h]
 \makebox[\textwidth][c]{
 \hspace*{-0.5cm}\includegraphics[width=1.0\textwidth]{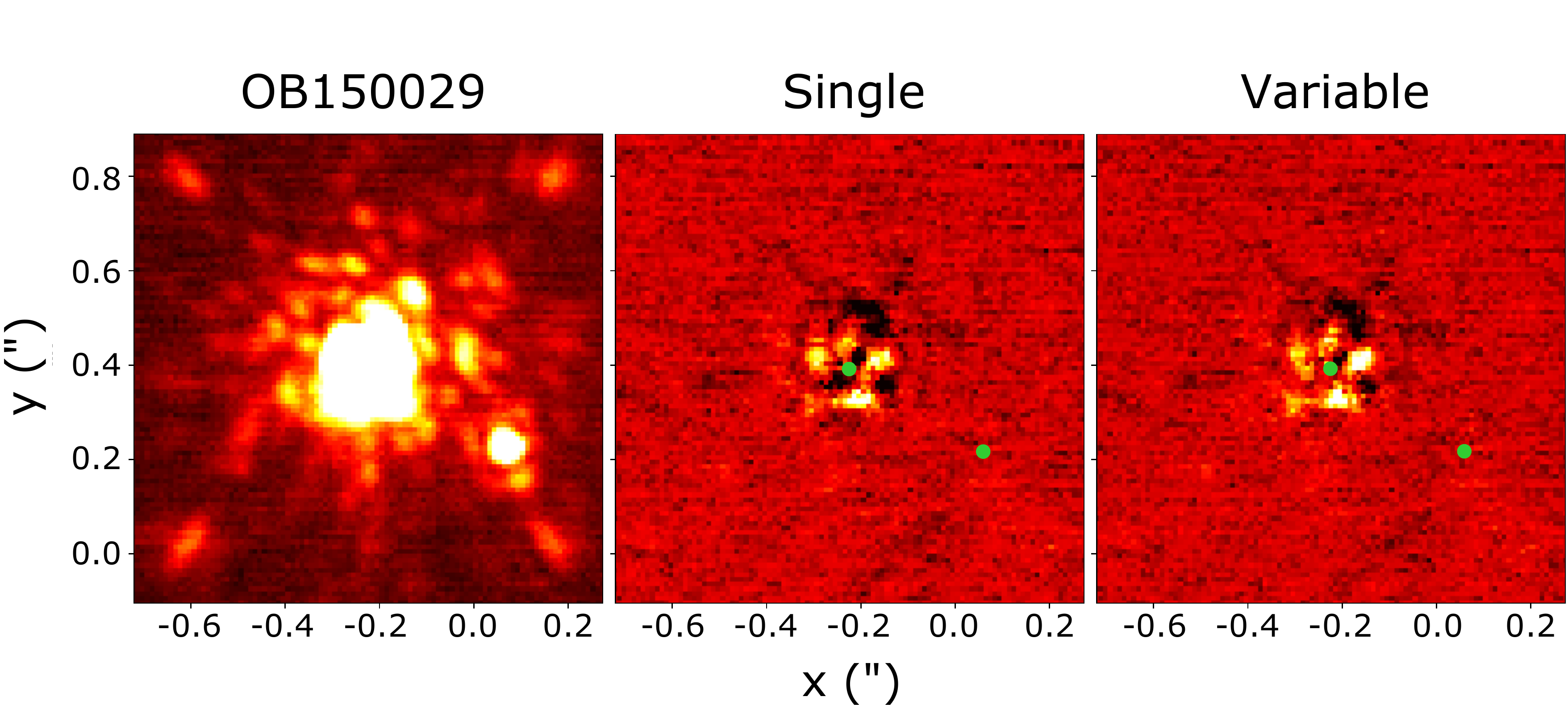}}
 \caption{\footnotesize Zoomed image of the target OB150029 (\textit{left}) and residual images from the single-PSF (\textit{middle}) and variable-PSF (\textit{right}) modes. Residual images have identical color scales, and green points show the location of star detections.} \label{fig:ob150029-targ-res}
\end{figure}
\begin{figure}[!h]
  \subfloat{\includegraphics[width=0.333\textwidth]{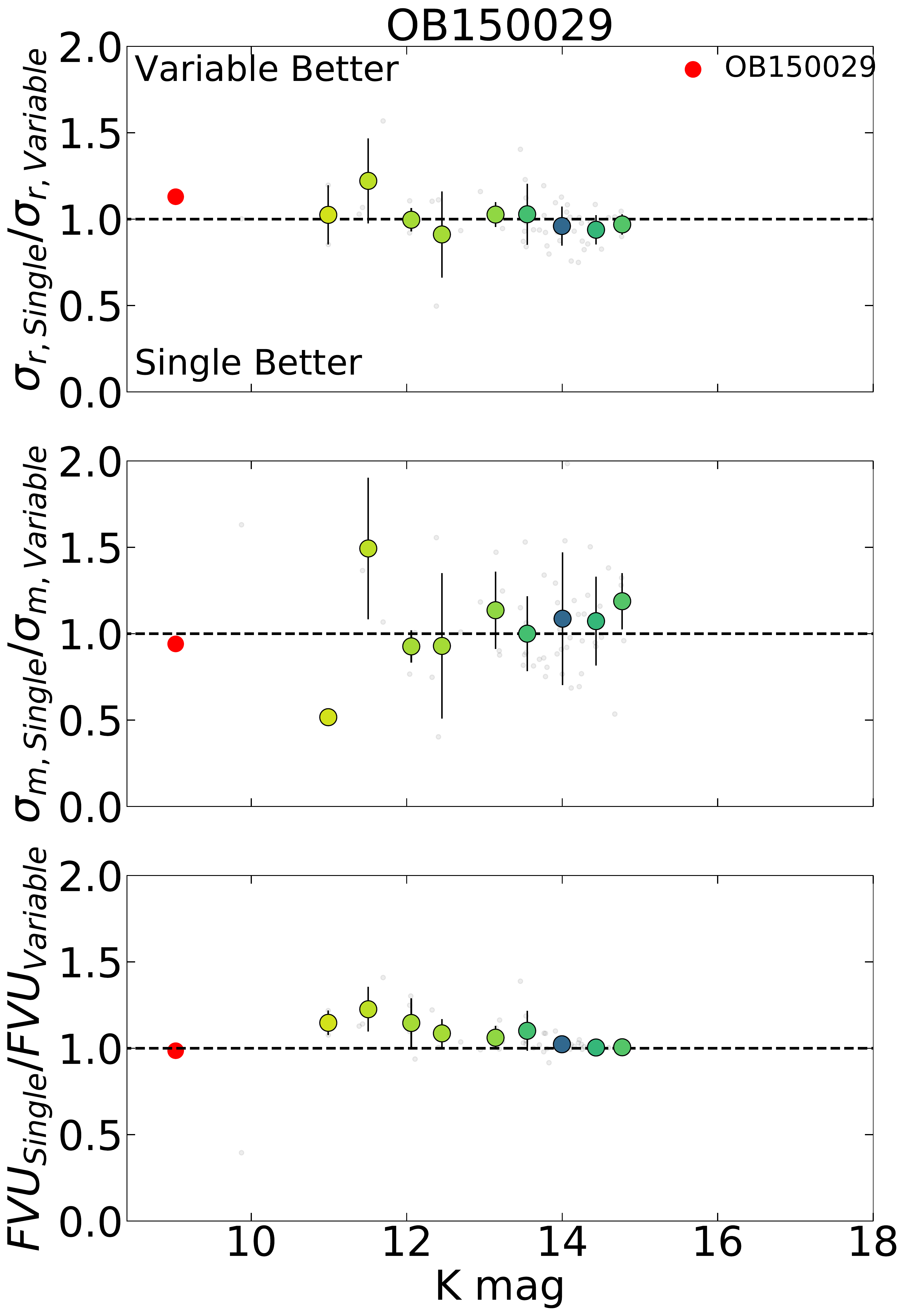}}
  \subfloat{\includegraphics[width=0.289\textwidth]{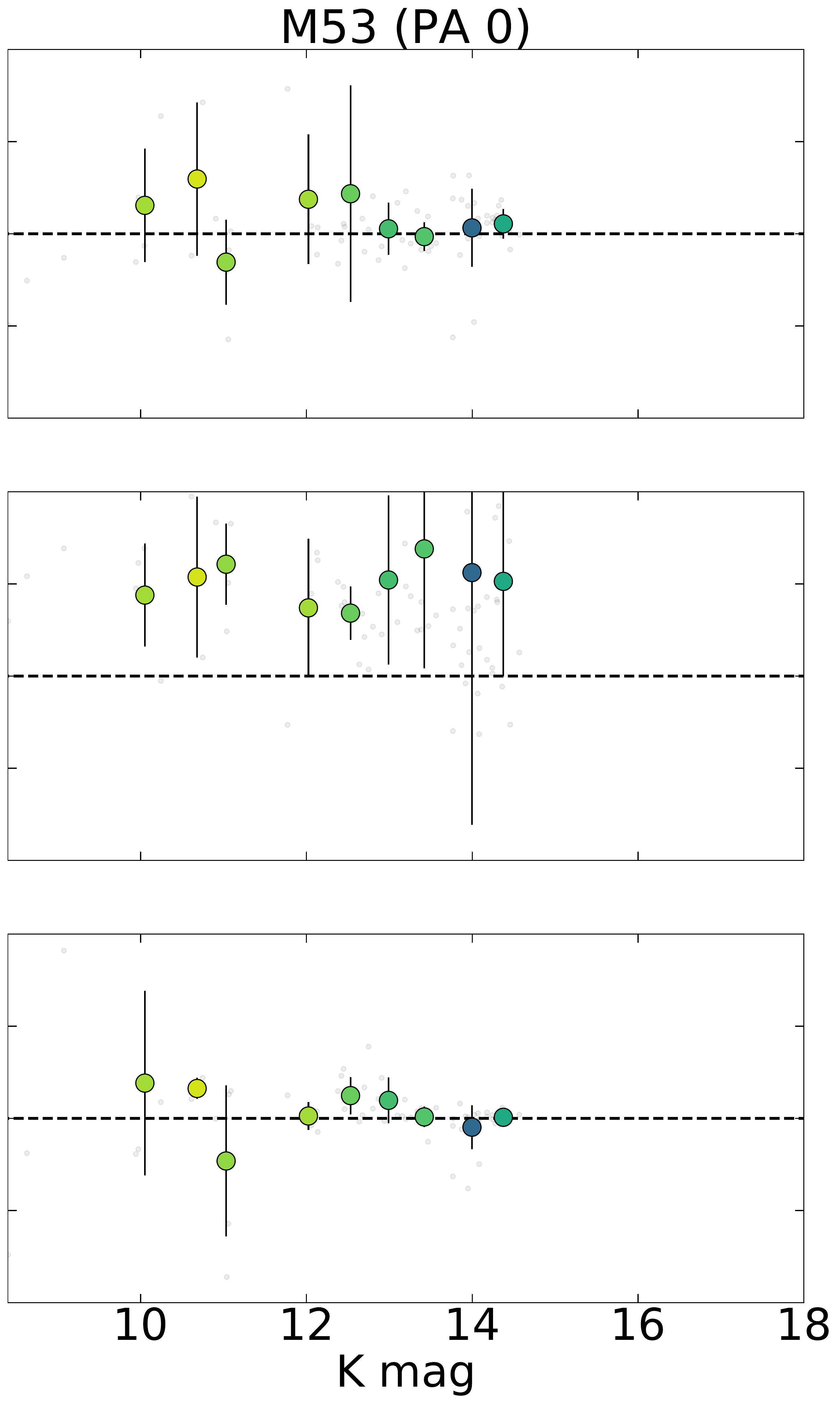}}
   \subfloat{\includegraphics[width=0.335\textwidth]{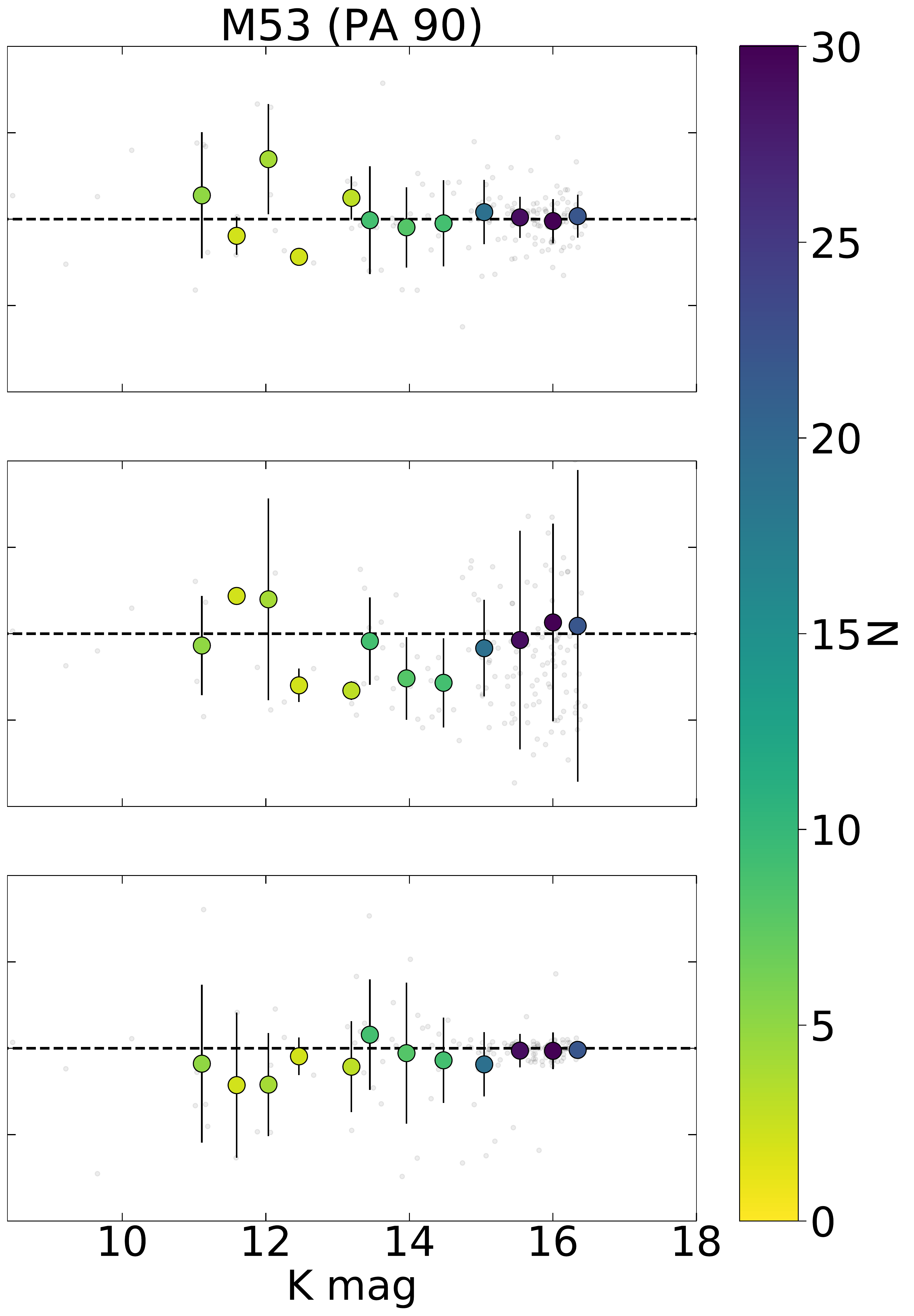}}
  \caption{\footnotesize Same as Figure \ref{fig:gc-astrom}, but for OB150029 (\textit{left column}), M53 PA 0 (\textit{middle column}), and M53 PA 90 (\textit{right column}).} \label{fig:ob-m53-astromfvu-ratios}
\end{figure}
\indent For the microlensing target itself, the AIROPA analysis shows a modest improvement of ${\sim}13\%$ in the astrometric precision in variable-PSF mode ($\sigma_{r}=0.223$ mas) over single-PSF mode ($\sigma_{r}=0.252$ mas). The upper-left panel of Figure \ref{fig:ob-m53-astromfvu-ratios} shows the ratio of astrometric residuals for all stars detected in both PSF modes (binned by magnitude), OB150029 is shown by the red data point. The photometric uncertainty and FVU evaluated on the target for both PSF modes are very similar, within 5\% for single-PSF and variable-PSF modes. It should not be surprising that the PSF fitting errors and metrics are quite similar between single-PSF and variable-PSF for the target star in this case since the location of the target on the detector is nearly on-axis (see left panel of Figure \ref{fig:ob150029-targ-res}) and the PSF shapes are not significantly different for the static PSF and the central PSF from the variable-PSF mode. The residual images show effectively identical over-subtracted and under-subtracted features for both PSF modes (middle and right panel of Figure \ref{fig:ob150029-targ-res}). We note there is an unrelated neighbor star approximately 0.36 arcseconds to the SW of the target, with a flux ratio of $f/f_{\textrm{OB150029}}\, {\sim}\, 0.1$. While the stellar profile of this fainter neighbor may have a small absolute effect on the SW wing of the OB150029 PSF, the magnitude of this effect will be the same for each case. Additionally, inspection of the residuals at the location of the nearby neighbor show no statistically significant signal above or below the background noise in both cases.
\\
\indent The left column of Figure \ref{fig:ob-m53-astromfvu-ratios} shows the relationship (i.e. ratio) between the fitting metrics as a function of magnitude, and Table \ref{tab:PhotAstromFVU-results} reports the results for each metric and PSF mode. Generally, the same trend appears in the microlensing dataset as the GC datasets; a marginal improvement in the precision and fitting residuals in variable-PSF mode compared to single-PSF mode. The largest improvement is seen in both astrometric precision and FVU for variable-PSF mode, ${\sim} 10\%$ smaller than the single-PSF mode results.
\\
\indent As a final test for this microlensing dataset, we performed an identical AIROPA analysis while omitting the target itself as a PSF reference star. Ideally, the results of such a test will reveal the influence (if any) that the target (i.e. brightest star in the field) has on the PSF modeling and extraction as well as the fitting parameters and residuals measured by the single-PSF and variable-PSF algorithms. The results of this re-analysis show indistinguishable results for the astrometric and photometric precision for the microlensing target ($\sigma_{r}=0.223$ mas, $\sigma_{m}=0.025$ mag), as well as an FVU value that is consistent (within 1$\sigma$) of the value measured while including the target as a PSF reference star.

\subsection{M53} \label{sec:results-m53}
As described in Section \ref{sec:m53-data}, the two M53 datasets have significantly different data qualities. In particular, the PA 0$^\circ$ dataset has the largest variance in SR, FWHM, and RMS WFE of all data in this study. One cause of this is the observing window for the PA 0$^\circ$ data spans a time frame where the DIMM and MASS monitors recorded a particularly high seeing measurement (right panel of Figure \ref{fig:dimmmass_datasets}, green-colored data). This is likely due to high cirrus cloud cover or other turbulent layers in the high-altitude atmosphere during two of the PA 0$^\circ$ exposures. A relatively small total number of PA 0$^\circ$ images were taken on the night (see Table 1 of \cite{service:2016a}), and the frames that correspond to the DIMM/MASS spike seen in Figure \ref{fig:dimmmass_datasets} happen to be the images that have the most observed sources in common with with the PA 90$^\circ$ images that were taken approximately one hour later in the same night. Because we want to maintain consistency amongst all datasets and to maximize the total number of cross-referenced stars in both PAs, we chose to keep all of the PA 0$^\circ$ frames for this particular dither position. This mostly affects the fainter stars in the field, as can be seen by comparing the limiting magnitudes in the PA 0$^\circ$ and PA 90$^\circ$ panels in Figure \ref{fig:ob-m53-astromfvu-ratios}. We note that AIROPA (and the Arroyo software) associates two PA 0$^\circ$ images with the large DIMM and MASS measurement.

\begin{table}[!h]
\caption{Average photometric, astrometric uncertainties, and FVU results for each dataset}
\setlength{\tabcolsep}{7.0pt}
\begin{center}
\begin{tabular}{lccccccc}
    \hline\hline
    {} & {} & \multicolumn{2}{c}{$\overline{\sigma}_{m}$ [mag]} & \multicolumn{2}{c}{$\overline{\sigma}_{r}$ [mas]} & \multicolumn{2}{c}{$\langle FVU \rangle$}\\
    \cmidrule(lr){3-4}\cmidrule(l){5-6}\cmidrule(l){7-8}
        Field & PSF Mode & All & $m_{K} < 13$ & All & $m_{K} < 13$ & All & $m_{K} < 13$\\
        \hline
        \multirow{1}{*}{GC (GQ)} & Single & $0.036$ & $0.018$ & $0.815$ & $0.472$ & $0.047 \pm 0.070$ & $0.005 \pm 0.017$\\
        {} & Variable & $0.036$ & $0.019$ & $0.806$ &$0.427$ & $0.047 \pm 0.070$ & $0.005 \pm 0.017$\\
        \hline
        \multirow{1}{*}{GC (AQ)} & Single & $0.040$ & $0.021$ & $1.111$ & $0.732$ & $0.050 \pm 0.062$ & $0.011 \pm 0.030$\\
        {} & Variable & $0.039$ & $0.021$ & $1.069$ & $0.711$ & $0.045 \pm 0.057$ & $0.008 \pm 0.022$\\
        \hline
        \multirow{1}{*}{GC (PQ)} & Single & $0.047$ & $0.033$ & $1.350$ & $1.012$ & $0.046 \pm 0.052$ & $0.012 \pm 0.018$\\
        {} & Variable & $0.042$ & $0.028$ & $1.327$ & $0.962$ & $0.045 \pm 0.051$ & $0.011 \pm 0.018$\\
        \hline
        \multirow{1}{*}{OB150029} & Single & $0.037$ & $0.032$ & $0.613$ & $0.454$ & $0.105 \pm 0.079$ & $0.021 \pm 0.021$\\
        {} & Variable & $0.035$ & $0.031$ & $0.622$ & $0.424$ & $0.101 \pm 0.078$ & $0.019 \pm 0.019$\\
        \hline
        \multirow{1}{*}{M53 (PA 0)} & Single & $0.068$ &  $0.063$ & $1.071$ & $0.670$ & $0.162 \pm 0.147$ & $0.037 \pm 0.059$\\
        {} & Variable & $0.049$ &  $0.045$ & $1.040$ & $0.626$ & $0.167 \pm 0.156$ & $0.036 \pm 0.057$\\
        \hline
        \multirow{1}{*}{M53 (PA 90)} & Single & $0.051$ & $0.036$ & $0.962$ & $0.306$ & $0.202 \pm 0.162$ & $0.002 \pm 0.001$\\
        {} & Variable & $0.059$ & $0.039$ & $0.972$ & $0.307$ & $0.215 \pm 0.176$ & $0.002 \pm 0.001$\\\hline
\end{tabular}
\end{center}
{\raggedright \footnotesize{\textbf{Note}. Average photometric uncertainty in magnitudes is given by $\overline{\sigma}_{m}$ (columns 3-4), average astrometric uncertainty in milliarcseconds is given by $\overline{\sigma}_{r}$ (columns 5-6). The mean FVU and 1$\sigma$ error is given in columns 7-8.}\par}
\label{tab:PhotAstromFVU-results}
\end{table}

\begin{figure}[!h]
  \centering
  \subfloat{\includegraphics[width=0.95\textwidth]{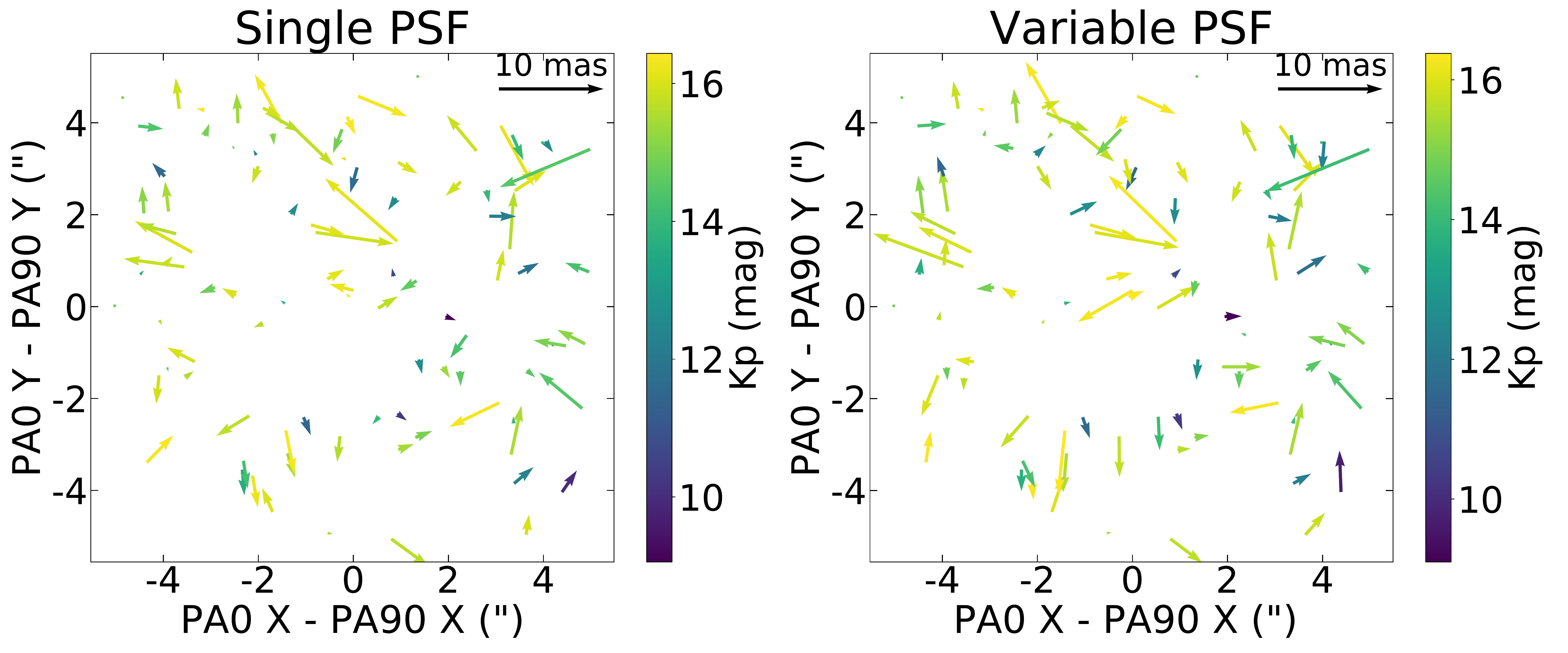}}\\
  \hspace{-1cm}
  \subfloat{\includegraphics[width=0.95\textwidth]{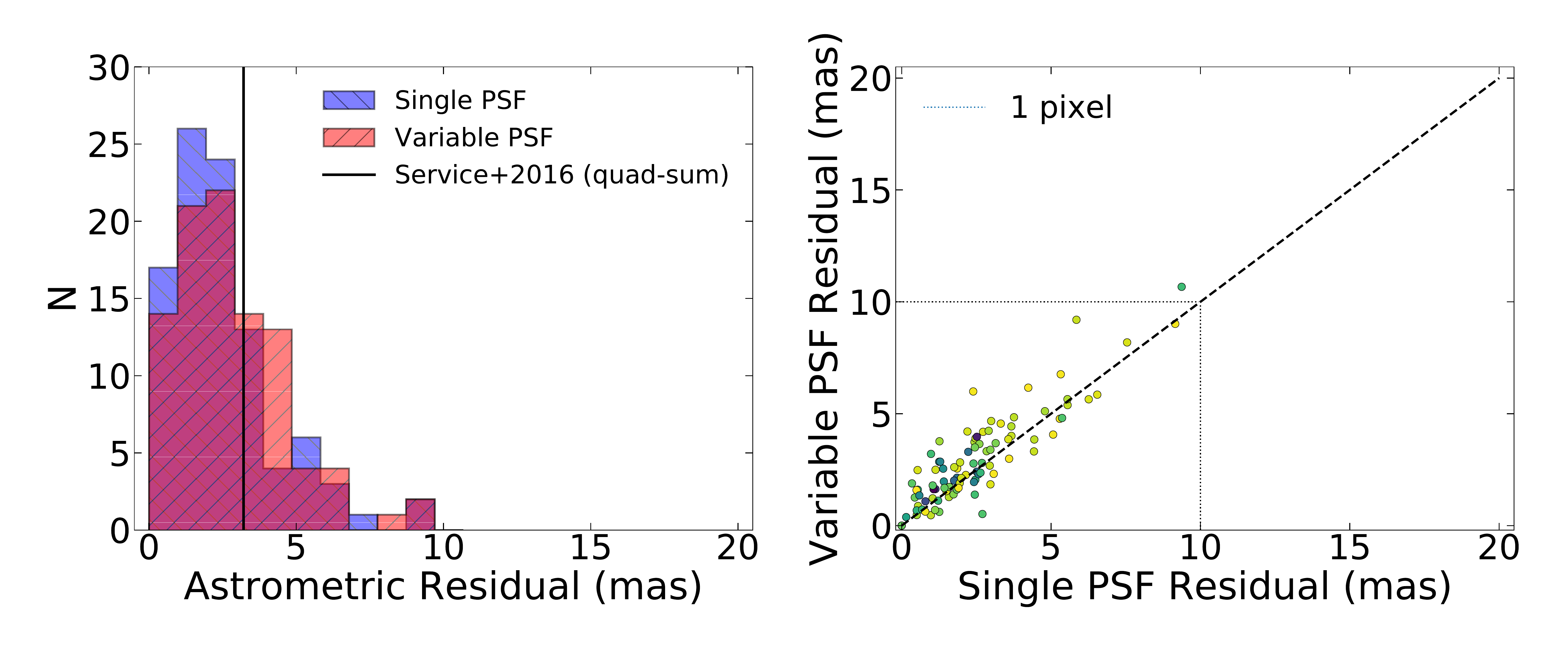}}
  \caption{\textit{Upper-left}: Quiver plot showing single-PSF astrometric residuals from the PA$=$0 $\rightarrow$ PA$=$90 transformation. \textit{Upper-right}: Same as upper-left, but for variable-PSF mode. \textit{lower-left}: Distribution of astrometric residuals from both PSF mode transformations. \textit{lower-right}: Correlation between astrometric residuals for both PSF modes, colored by Kp magnitude.} \label{fig:m53_PA_compare}
\end{figure}

\subsubsection{M53 PA 0} \label{sec:result-m53-pa0}
A total of 98 stars were cross-matched across at least five of the six PA 0$^\circ$ frames, with 10 of the stars used to build the initial PSF model. The location and brightness of the selected PSF reference stars are given by the colored data points in the bottom panels of Figure \ref{fig:m53_image_grid}. The difference in astrometry, photometry, and FVU between single-PSF and variable-PSF modes largely follow the same trends seen in the GC and OB150029 analyses (Tables \ref{tab:PhotAstromDiff-results} and \ref{tab:PhotAstromFVU-results}). Interestingly, the most significant difference for the M53 PA 0$^\circ$ single and variable PSF modes comes from the comparison of the photometric uncertainties (middle panel of Figure \ref{fig:ob-m53-astromfvu-ratios}). The results show that the variable-PSF mode gives significantly lower photometric errors than the single-PSF mode (Table \ref{tab:PhotAstromFVU-results} columns 3$-$4 for M53 PA 0$^\circ$). \\
\indent It is worth noting a similar, yet less significant trend in photometric precision is also seen in the poorest quality GC data set (GC PQ). These trends suggest that for lower-quality, less uniform point sources, the spatially variable PSF model more precisely measures the intensity weighted integrated coordinate of the stars compared to the static PSF model. Additionally, the magnitudes and colors of the M53 stars as a function of off-axis location show a significant improvement in variable-PSF mode when comparing to an absolute \textit{HST} reference (Section \ref{sec:m53-hst}).

\subsubsection{M53 PA 90} \label{sec:result-m53-pa90}
The higher quality PA 90$^\circ$ data has a total of 160 stars cross-matched in the images, with the same 10 stars used as PSF reference stars (lower-right panel of Figure \ref{fig:m53_image_grid}). Like the previous data sets, there is little to no improvement in the variable-PSF fitting metrics compared to the single-PSF results. The difference in stellar positions and magnitudes also follow similar trends seen in the previous data (Table \ref{tab:PhotAstromDiff-results}). The average photometric error for all stars is ${\sim}15$\% larger in the variable mode compared to single mode. The average astrometric error is nearly identical in both modes, at a ${\sim}1$\% difference. The FVU value in both modes are within $1\sigma$ of each other (bottom row of Table \ref{tab:PhotAstromFVU-results}). \\
\indent Finally, we investigated the difference between astrometric residuals in both PSF modes after matching and transforming the PA 0$^\circ$ star list to the PA 90$^\circ$ list (and vice versa). We perform a first-order transformation using both stacked star lists from each PA as the master reference list and star list to be transformed in both cases. If the variable-PSF mode performs better, then we would expect astrometric residuals from this transformation process to be smaller for the spatially varying PSF mode, particularly for larger off-axis stars in the frames. Figure \ref{fig:m53_PA_compare} shows the resulting astrometric quiver plots in the top panels, and the distribution and correlation between the astrometric residuals in both PSF modes in the bottom panels. The two distributions are very similar, and correlations largely follow the line of unity with a dispersion of ${\sim}$a few mas (bottom right panel of Figure \ref{fig:m53_PA_compare}), which show that the astrometric residuals from the variable-PSF mode are not measurably smaller than the single-PSF residuals.

\subsubsection{HST Cross Reference} \label{sec:m53-hst}

\begin{figure}[!h]
  \centering
  \subfloat{\includegraphics[width=0.96\textwidth]{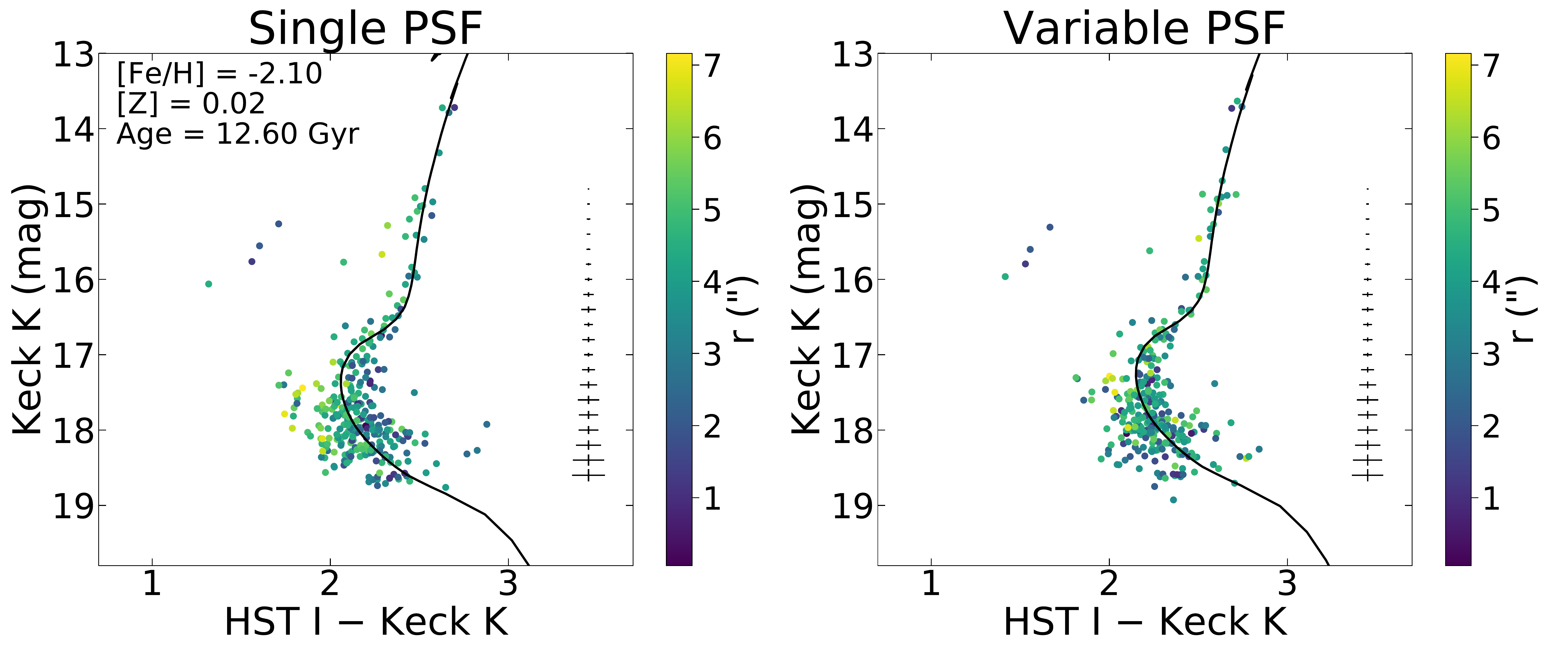}}\\
  \hspace{-1.2cm}
  \subfloat{\includegraphics[width=0.93\textwidth]{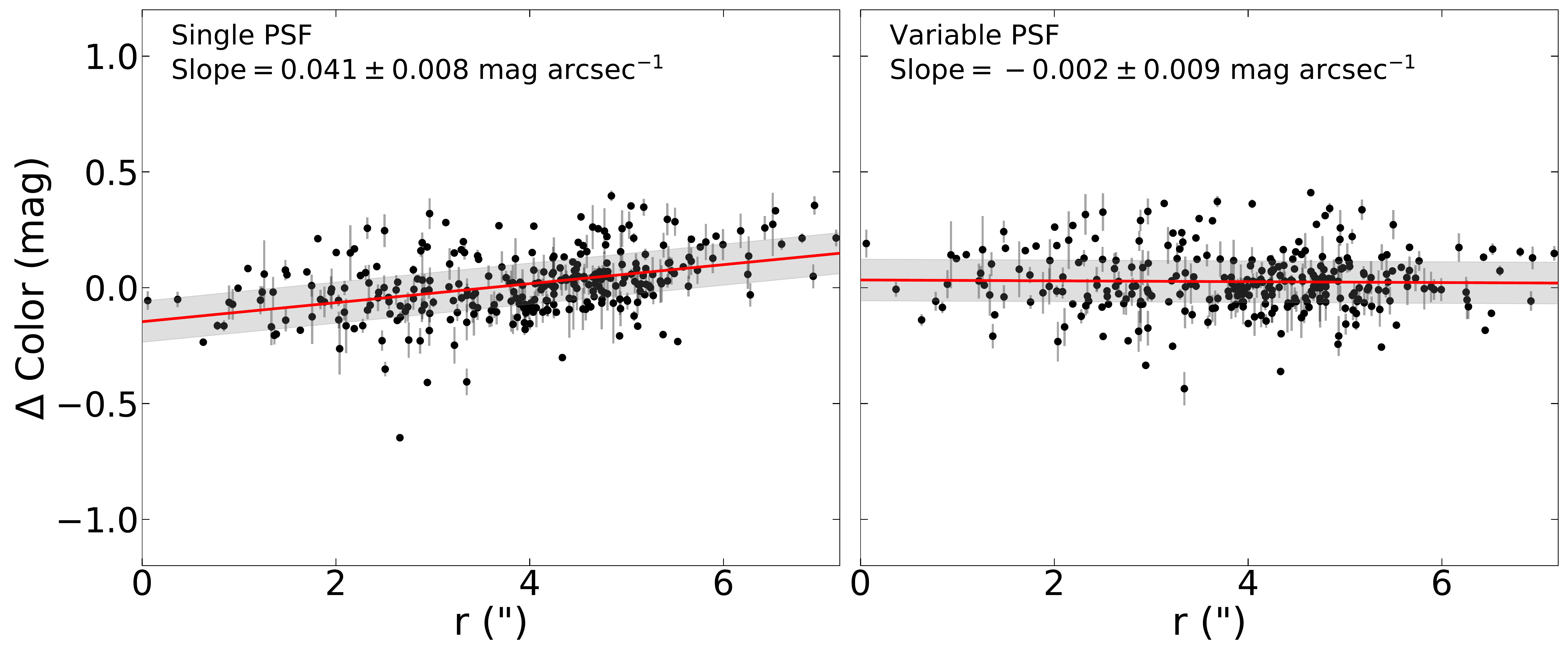}}
  \caption{\textit{Top row}: The M53 \textit{HST I}$-$Keck $K$ vs. Keck $K$ CMDs for both single-PSF and variable-PSF star lists. The solid black curve gives the best-fit SPISEA isochrone, with age, metallicity, mass fraction, and extinction from \cite{dotter:2011a,wagner:2016a,schlafly:2016a}. \textit{Bottom row}: The measured color index difference (i.e. empirical $-$ isochrone) in single-PSF and variable-PSF modes as a function of off-axis radial distance. The solid red line shows the linear best-fit to each distribution, with the shaded grey region showing the 1$\sigma$ spread around the best-fit. The distribution of color differences for single-PSF mode has a clear trend in radial distance, while the distribution is largely constant for variable-PSF mode.} \label{fig:m53_cmd_colorshift}
\end{figure}

\indent Both PA's star lists were independently transformed to an \textit{HST} reference frame described in Section this section, with the same intention of examining the difference in astrometric residuals between single and variable modes. Figure \ref{fig:m53_PA_compare_hst} shows the PA 90$^\circ$ astrometric residual quiver plots in the upper panels, and corresponding distribution of astrometric residuals as well as the correlation between single-PSF and variable-PSF residuals in the lower panels. The distribution of astrometric residuals peaks at $1.52$ mas in both PSF modes, this is consistent with the residual distortion on the NIRC2 detector that was found in \cite{service:2016a} (denoted by the vertical black line in the bottom left panel of Figure \ref{fig:m53_PA_compare_hst}). The weighted mean and central 68\% of both PSF mode distributions is given in Table \ref{tab:m53-AstromResidual-results}. This table includes all variations of star list transformations that were performed on the M53 data sets (i.e. NIRC2, \textit{HST}, and \textit{Gaia} described below).
\\
\indent Further, we used the HST photometric measurements as an absolute standard to compare the NIRC2 single-PSF and variable-PSF photometry against. We do this by generating a color-magnitude diagram (CMD) for each PSF mode and PA, where we compare the HST I $-$ Keck K color against the Keck K magnitudes for all cross-matched stars in each catalog. The NIRC2 data were photometrically calibrated to 2MASS Ks band magnitudes for this process. 
We cross-identify two bright stars in common between the NIRC2 data and 2MASS (2MASS J13125435+1810172 and 2MASS J13125408+1810194). From this cross-identification we calculate a zero-point offset of 3.70 and apply this to all stars in each catalog, including a calibration error of 0.06 magnitudes.
\\
\indent We then used the open-source package SPISEA \cite{hosek:2020a} to generate a synthetic population for the cluster. We used parameters for the cluster age, metallicity, and mass fraction from \cite{dotter:2011a} and \cite{wagner:2016a}. We also adopted the extinction law of \cite{schlafly:2016a} when generating the population. The top panels of Figure \ref{fig:m53_cmd_colorshift} show the single-PSF and variable-PSF CMDs. The data are colored by off-axis radial distance in the NIRC2 frames, and the best-fit isochrone from SPISEA is overlaid (black curve). Characteristic errors in color and magnitude for the CMDs are given by the vertical columns. There is a clear trend seen in the comparison between the CMD stars and the isochrone model in the single-PSF CMD. The difference in color between the stars and isochrone model shows a relatively strong dependence on radial distance in the NIRC2 frames. This same dependence is not seen in the comparison of the variable-PSF CMD and best-fit isochrone model. The lower panels of Figure \ref{fig:m53_cmd_colorshift} show the color difference measurements (and errors) as a function of off-axis distance. A linear fit is made to each distribution, and the resulting slopes are $0.041 \pm 0.008$ mag arcsec$^{-1}$ for single-PSF mode and $-0.002 \pm 0.009$ mag arcsec$^{-1}$ for variable-PSF mode. These results suggest roughly an order of magnitude improvement in the photometric performance of the variable-PSF mode compared to single-PSF mode. This marks the first definitive example of a significant improvement shown by variable-PSF mode over the single-PSF mode for on-sky science data.


\clearpage

\subsubsection{Gaia Cross Reference} \label{sec:m53-gaia}

\indent We continue our comparison of the single-PSF and variable-PSF catalogs by transforming both lists to the Gaia EDR3 catalog \cite{brown:2021a}. While Gaia has delivered exquisite astrometric data for over one billion sources now, there are a sizable fraction of spurious values reported by the instrument that are not so reliable. Recently, it has been shown that numerous astrometric solutions reported by Gaia suffer quite significantly for low signal-to-noise sources as well as for sources in crowded fields \cite{rybizki:2022a}. Given that the central region of M53 is crowded, we perform a careful investigation of Gaia astrometric quality flags for the stars cross-matched between NIRC2 and Gaia. A consideration that we make when matching the two PA's to Gaia are the astrometric fitting errors as described in \cite{brown:2021a} and \cite{brown:2018a}. In particular the ``astrometric excess noise ($\epsilon$)" parameter is described as the excess uncertainty that must be added in quadrature to obtain a statistically acceptable astrometric solution. Further, the excess term $\epsilon$ is introduced in order to effectively reduce the statistical weight of observations that may be affected by things like instrument and attitude modeling errors. Another error term we consider is the ``astrometric excess noise significance", which is described as a dimensionless measure of the significance of the calculated $\epsilon$. A value of $> 2$ indicates that the reported $\epsilon$ is ``probably significant" \cite{lindegren:2012a}. Further details on these astrometric error terms can be found in \cite{lindegren:2012a}.
\\
\begin{figure}[!h]
 \centering
 \includegraphics[width=0.9\textwidth]{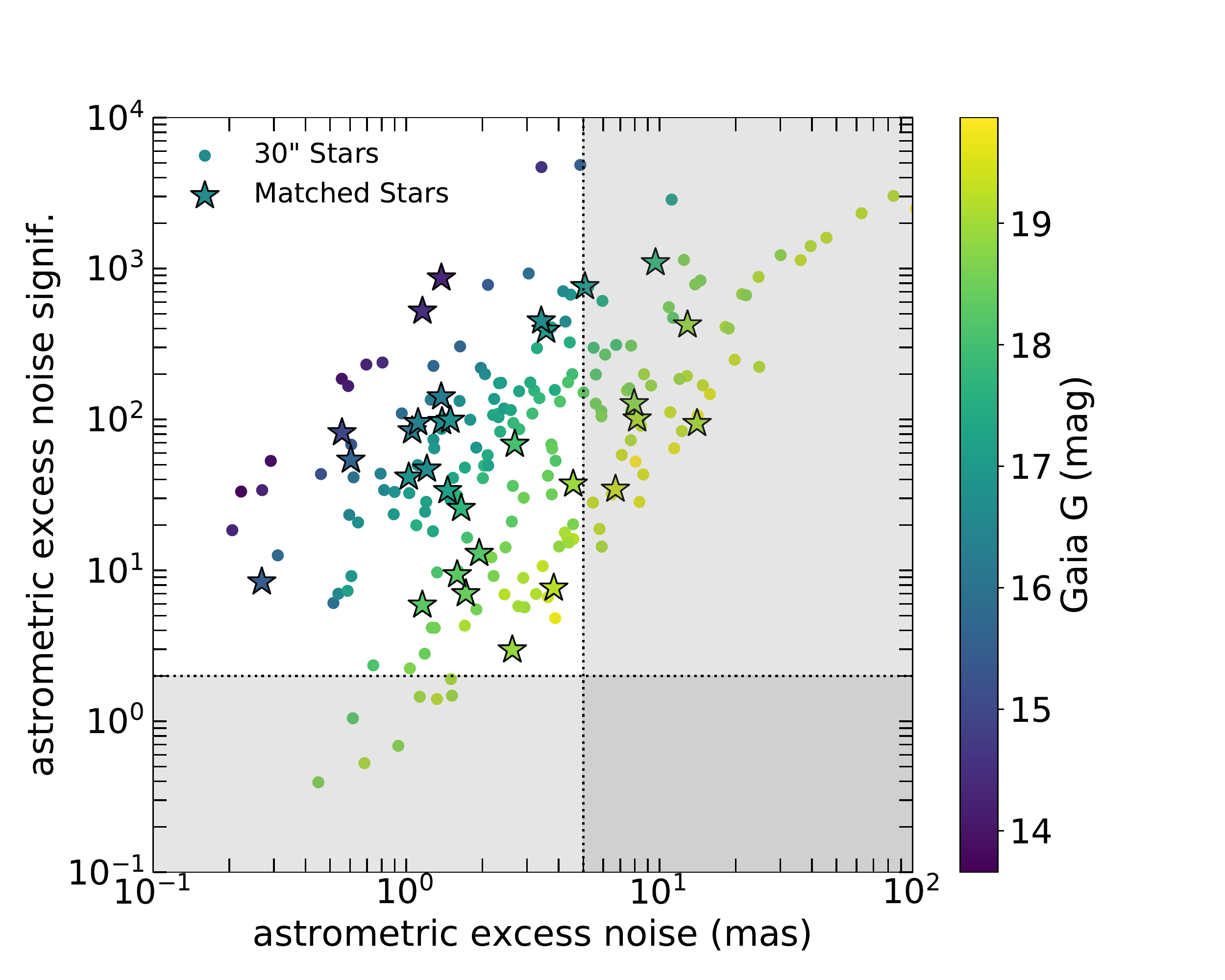}
 \caption{\footnotesize Astrometric excess noise significance vs. astrometric excess noise as a function of Gaia $G$ magnitude in the 30 arcsecond field surrounding the center of the NIRC2 images. The reference stars used for the NIRC2 $\rightarrow$ Gaia transformation are marked as star symbols. Dotted vertical and horizontal lines denote the noise and noise significance thresholds used to cut reference stars.\label{fig:m53_aen}}
\end{figure}
\indent Figure \ref{fig:m53_aen} shows the values reported by Gaia for the two astrometric error terms described above. The star symbols in the figure represent 32 NIRC2 stars that were cross-identified in Gaia. All other Gaia stars within a 30 arcsecond radius of the center of the NIRC2 frame are plotted as well, and both distributions are colored by Gaia G magnitude. Given the higher astrometric precision in both the Keck and \textit{HST} data, we choose to make a cut on the Gaia stars at $\epsilon < 5$ mas and significance $< 2$ (shaded regions in Figure \ref{fig:m53_aen}), which leaves 24 Gaia stars used for the final transformation. As a note, we tested more strict and less strict cutting thresholds for this process (i.e. some stars rejected by their \textit{significance} value) and found similar results for each scenario. The cut at $\epsilon <  5$ mas left a sufficient number of stars needed for a reliable first-order transformation, while rejecting likely astrometric outliers (which are mostly faint, low-SNR stars).
\\
\begin{figure}[!h]
  \centering
  \subfloat{\includegraphics[width=0.95\textwidth]{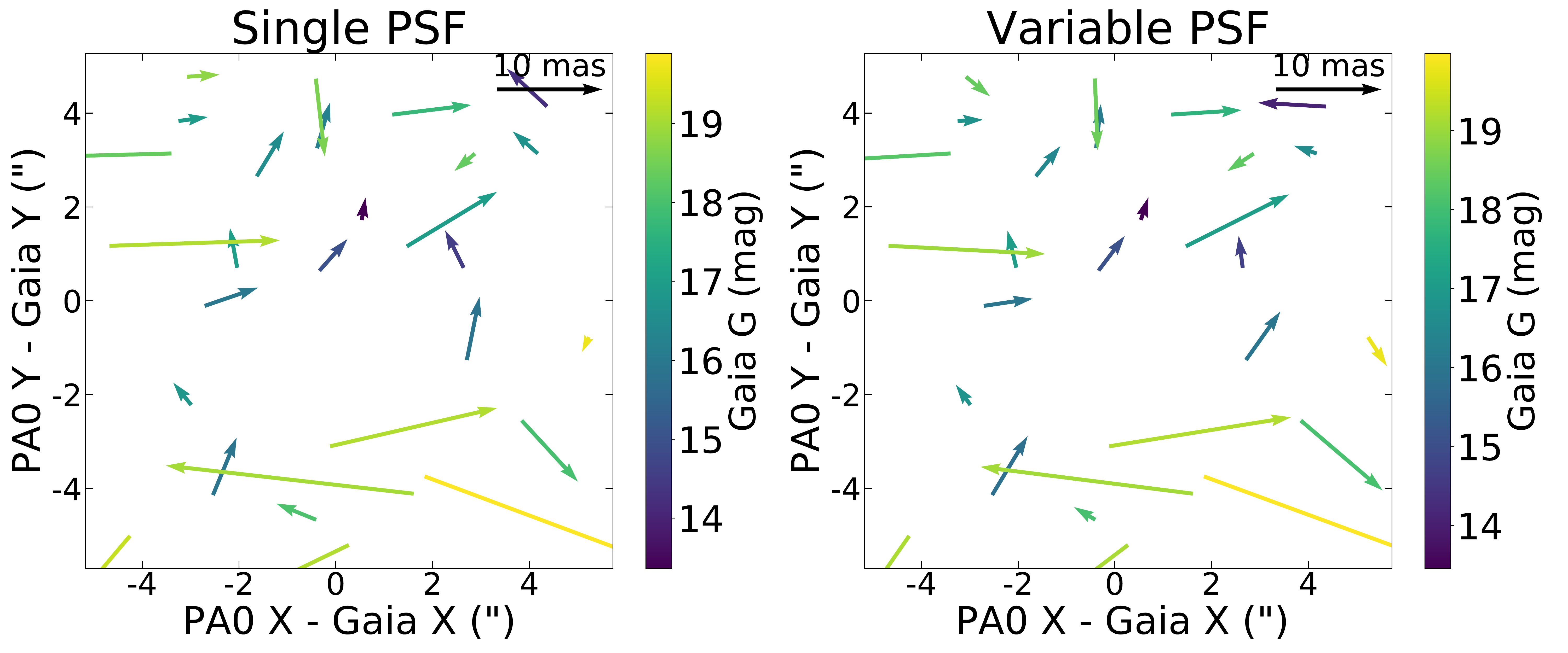}}\\
  \hspace{-1cm}
  \subfloat{\includegraphics[width=0.95\textwidth]{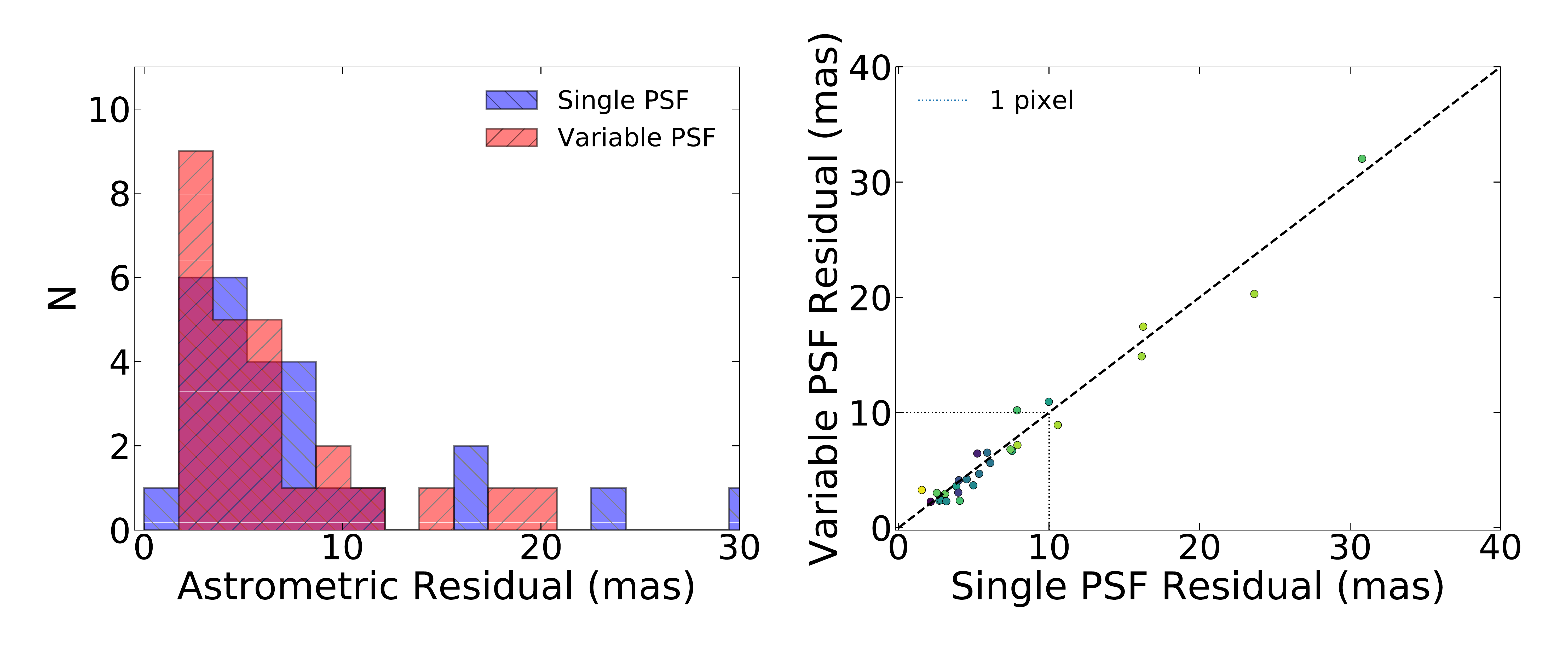}}
  \caption{Same as Figures \ref{fig:m53_PA_compare} and \ref{fig:m53_PA_compare_hst} but for stars transformed to Gaia EDR3. The color bar is given in Gaia G magnitude.} \label{fig:m53_PA_compare_gaia}
\end{figure}
\indent The quiver plots in the top panels of Figure \ref{fig:m53_PA_compare_gaia} show the 2D astrometric residuals after transforming the PA 0$^\circ$ stars to the Gaia reference frame. The distribution of astrometric residuals for both PA's peaks at $\sigma_{r}{\sim}5$ mas, and the faintest stars have the largest residuals, as expected. Table \ref{tab:m53-AstromResidual-results} shows the weighted mean for each distribution, where the mean and errors are largest for Gaia. As described earlier, there are several reasons why the Gaia residuals are largest - generally the Gaia astrometric precision is worse than \textit{HST} (and NIRC2), especially for crowded field sources. The astrometric excess noise and other error terms contribute overall larger residuals when transforming NIRC2 into the Gaia reference frame. Finally, the calculation of the weighted mean and central 68\% of the distribution is more significantly affected by the spread in residuals (and possible outliers) for the much smaller Gaia data set (24 stars).


\begin{table}[!h]
\caption{M53 astrometric residuals from NIRC2, HST, and Gaia}
\setlength{\tabcolsep}{11.0pt}
\begin{center}
\begin{tabular}{lccc}
    \hline\hline
    {} & {} & \multicolumn{1}{c}{PA 0}\\
    \cmidrule(lr){2-4}
        Ref. Frame & $N_{\textrm{matched}}$ & $\langle \sigma_{r_{\textrm{s}}}\rangle$ [mas] & $\langle \sigma_{r_{\textrm{v}}}\rangle$ [mas]\\
        \hline
        PA 90 & 94 & $2.53 ^ {+1.59}_{-0.47}$ & $2.82 ^ {+1.67}_{-0.51}$\\
        HST & 90 &  $1.93 ^ {+1.39}_{-0.44}$ & $2.04 ^ {+1.43}_{-0.45}$\\
        Gaia & 24 &  $7.61 ^ {+2.76}_{-1.19}$ & $7.35 ^ {+2.71}_{-1.12}$\\
        \hline
        {} & {} & PA 90 & {}\\
        \cmidrule(lr){2-4}
        PA 0 & 94 & $2.43 ^ {+1.56}_{-0.49}$ & $2.89 ^ {+1.70}_{-0.68}$\\
        HST & 157 &  $1.85 ^ {+1.36}_{-0.43}$ & $1.99 ^ {+1.42}_{-0.45}$\\
        Gaia & 24 &  $7.39 ^ {+2.72}_{-1.20}$ & $7.80 ^ {+2.79}_{-1.25}$\\
        \hline
\end{tabular}
\end{center}
\label{tab:m53-AstromResidual-results}
\end{table}

\section{Discussion and Conclusion} \label{sec:conclusion}

\indent We have analyzed several on-sky NIRC2 datasets with the the single PSF mode and spatially variable PSF mode in AIROPA. We find that the performance of AIROPA is reliable across different conditions including poor, average, and good quality seeing, crowded or sparse stellar fields, varying numbers and brightnesses of PSF reference stars, and in different telescope PA's.
\\
\indent Our analysis of the datasets largely show only a marginal improvement in photometric or astrometric residuals between the static PSF model and spatially variable PSF model within AIROPA. A comparison of the FVU metric between the two modes show similar results. This implies that the ability of AIROPA to reconstruct the PSF for a wide range of on-sky data remains limited by unaccounted for static or quasi-static aberrations in the telescope. We also show that the effect of varying atmospheric conditions, number and spatial location of selected PSF reference stars, and telescope PA do not have a significant effect on the performance of AIROPA in either of the modes.
\\
\indent One metric that does show a significant improvement in the variable-PSF mode over the single-PSF mode is a comparison of the color spread of M53 stars in a Keck + \textit{HST} CMD. We find a measurable trend in photometry as a function of off-axis location in the single-PSF mode star catalog, whereas the variable-PSF mode star catalog shows no trend. After comparing the fitted trend line for each case, we find that the photometric performance is improved by ${\sim}10\times$ for the spatially variable PSF mode. This represents the first significant improvement from the variable-PSF mode that has been found for on-sky science data.
\\
\indent Comparing the FVU metrics between PSF fitting modes across all datasets shows at best a ${\sim}5\%$ improvement in bright stars for the spatially varying PSF model over a static PSF model. This is significantly less than what has been shown in tests on simulated GC data \cite{turri:2022}. For OB150029 we measure an astrometric precision that is ${\sim}13\%$ smaller for the variable-PSF mode over the single-PSF mode. However, the FVU metric for this target does not follow the same improvement seen in the astrometry. For the M53 data, we find that the astrometric residuals between single and variable-PSF modes are quite similar when transforming a stack of PA $=$ 0$^\circ$ frames onto the PA $=$ 90$^\circ$ reference frame. Further, when transforming the two PA's to the Gaia and \textit{HST} reference, we find comparable astrometric residuals in both PSF modes.
\\
\indent Finally, for most of the fitting metrics (except for the Keck + \textit{HST} photometric comparison), we largely confirm the result of \cite{turri:2022} which shows no significant improvement in fitting residuals for the spatially variable PSF mode in AIROPA with on-sky data. It is hypothesized that there remains static or quasi-static instrumental aberrations that persist in the telescope and are not being fully characterized by afternoon phase-diversity measurements. This hypothesis is backed up by a recent analysis that shows a dominant source of error in the PSF comes from primary segment misalignments (O. Beltramo-Martin in private communication). The work shows several hundred nm of wavefront error (WFE) coming from the primary piston segments, which can become misaligned as quickly as hours after initial alignment. Currently, the Keck-II primary mirror segments are realigned every two-weeks, however these results suggest the cadence may need to be increased in order to minimize any contribution to the PSF error from this primary segment phasing. Future on-sky phase-diversity measurements should help in identifying the source(s) of instrumental aberrations that are not currently accounted for in fiber phase-diversity measurements. 

\acknowledgments

The data presented herein were obtained at the W. M. Keck Observatory, which is operated as a scientific partnership among the California Institute of Technology, the University of California and the National Aeronautics and Space Administration.
The Observatory was made possible by the generous financial support of the W. M. Keck Foundation.
The authors wish to recognize and acknowledge the very significant cultural role and reverence that the summit of Maunakea has always had within the indigenous Hawaiian community.
We are most fortunate to have the opportunity to conduct observations from this mountain.
\\
\indent SKT acknowledges support from the NSF through grant AST-1836016. We acknowledge support from the W. M. Keck Foundation, the Heising-Simons Foundation, the Gordon and Betty Moore Foundation, and the NSF (AST-1412615, AST-1518273). SKT thanks Taylor Weiss and Jeremy Van Kosh for helpful software discussions and comments on this manuscript. We thank the staff of Keck Observatory for their help with obtaining calibration data and overall support for our program.

\bibliography{refs} 
\bibliographystyle{spiejour} 

\end{document}